\documentclass[12pt]{JHEP}
\usepackage{epsfig}

\def\fun#1#2{\lower3.6pt\vbox{\baselineskip0pt\lineskip.9pt
  \ialign{$\mathsurround=0pt#1\hfil##\hfil$\crcr#2\crcr\sim\crcr}}}
\newcommand{\beq}{\begin{equation}}
\newcommand{\eeq}{\end{equation}}

\title{Lensing of ultra-high energy cosmic rays\\ in turbulent magnetic fields}

\author{Diego Harari$^a$, Silvia Mollerach$^b$, Esteban Roulet$^b$ and
Federico S\'anchez$^a$
\\$^a$Departamento de F\'\i sica, FCEyN, Universidad de Buenos Aires
\\Ciudad Universitaria - Pab. 1, 1428, Buenos Aires, Argentina
\\$^b$CONICET and Centro At\'omico Bariloche\\Av. Bustillo km 9.5, 8400, 
S.C. de Bariloche, Argentina
\\ Email: \email{harari@df.uba.ar, mollerac@cab.cnea.gov.ar, 
roulet@cab.cnea.gov.ar, federico@df.uba.ar}}

\abstract{We consider the propagation of ultra high energy cosmic rays 
through turbulent magnetic
fields and study the transition between
the regimes of single and multiple images of point-like
sources. The transition occurs at energies around
$E_c\simeq Z~41\ {\rm EeV}(B_{rms}/5\ \mu{\rm G})\left 
(L/ 2~{\rm kpc}\right)^{3/2}\sqrt{50~{\rm pc}/L_c}$,
where $L$ is the
distance traversed by the CR's with electric charge $Ze$ in the turbulent
magnetic field of root mean square strength $B_{rms}$ and coherence
length $L_c$. 
We find that 
above $2 E_c$ only sources
located in a fraction of a few \% of the sky can reach
large amplifications of its principal image or start
developing multiple images. New images appear in pairs with huge
magnifications, and they remain amplified over a  significant
 range of energies. At decreasing energies the fraction of the sky in
which sources can develop multiple images increases, reaching  about
50\% for  $E>E_c/2$. The magnification 
peaks become however increasingly narrower and for $E<E_c/3$  their
 integrated  effect
becomes less noticeable. If 
 a uniform magnetic field component is also present it would further narrow
down the peaks, shrinking the
energy range in which they can be relevant.
Below $E\simeq E_c/10$ some kind of scintillation
regime is reached, where many demagnified images of a source are present
but with
overall total magnification of order unity. 
We also search for lensing signatures in the AGASA data
studying two-dimensional correlations in angle and energy and find
some interesting hints.}

\keywords{High-energy cosmic rays}
\preprint{.}

\begin{document}

\section{Introduction}

One of the fundamental open problems in physics is to understand the
nature and origin of the ultra-high energy cosmic rays (UHECRs). The
increased aperture of the next generation of CR detectors (e.g. the
Auger Observatory) will be crucial to attempt to locate possible
astrophysical point-like sources from the observation of clustering or
anisotropies in the arrival directions. If the UHECRs are electrically
charged, as would be the case in the likely situation that they are
protons or heavier nuclei, a very important fact which has to be taken
into account is their deflection by the magnetic fields present along
their path. These deflections are expected to be quite large below the
ankle (i.e. for $E<5\times 10^{18}$~eV), and hence no CR `astronomy'
is possible below those energies since the information about the
original source direction cannot be recovered from the observed
arrival directions. On the other hand, since typical galactic magnetic
fields ($B\simeq {\rm few}\ \mu$G) are unable to confine CRs within
the Galaxy for energies above the
ankle, this supports the belief that UHECRs have an extragalactic
origin, in agreement with the absence of strong anisotropies toward
the Galactic disk being observed at those energies.
The decreasing magnetic deflections for increasing energies lend us
hope that a correlation among the arrival directions of different CR
events and between those and the
possible source directions will clearly show up with increased statistics
at energies $E\gg 10^{19}$~eV, when the CR trajectories  progressively 
approach the
rectilinear propagation regime.

The astrophysical magnetic fields \cite{ru88,kr94,be01} 
may be classified under two broad
categories: the regular and the random fields. Large scale regular
fields may be produced by adiabatic compression of preexisting cosmic
seed fields, by their amplification by  a dynamo mechanism or may be
related to the action of a galactic wind. They are known to exist in
our galaxy (and other spirals as well), where the field lines
essentially follow the spiral structure, with reversals in direction
taking place between neighboring arms.  The existence of coherent
fields in the halo or even larger scales is debatable, as is the
presence of regular fields on cluster scales.
The random component is present on galactic scales, probably
originating from supernova explosions which feed power on typical
scales of $\sim 100$~pc, which is latter transferred to smaller scales
due to the turbulence present in the high Reynold's number interstellar
medium. This leads to a Kolmogorov spectrum of magnetic field
fluctuations extending probably down to very small scales ($\sim
10^8$~m \cite{ar81}) and with an rms amplitude exceeding the typical
values of the uniform field by a factor $2\div 3$. In cluster's cores
$\mu$G fields have been observed, and their coherence length is
believed to be at the kpc scale. 
For the
intergalactic medium usually turbulent random fields are also assumed
to exist, but with much larger coherence lengths ($\sim 1$~Mpc). 
The amplitude of these fields is bounded from the
non-observation of Faraday rotation of distant polarized radio
sources, but it can be sizeable ($\sim 10^{-8}\div 10^{-7}$~G) if
confined in thick sheets e.g. around the supergalactic plane.

The existence of these magnetic fields will certainly affect the
observable properties of UHECRs (see e.g. 
\cite{gi80,be89,wa96,st97,me98,si99}).
As we have shown in previous works \cite{ha99,ha00}, besides inducing a change
in the arrival directions, the 
deflections caused by the magnetic fields can also lead to strong
lensing phenomena, including the formation of multiple images
and energy dependent magnifications or demagnifications of the CR
fluxes. In those works we dealt with the effects resulting from
the regular galactic magnetic field, which were shown to lead to
important consequences for $E/Z<30$~EeV (where $Ze$ is the assumed CR charge
and $1~{\rm EeV}\equiv 10^{18}$~eV). It is the purpose of the present
work to extend the analysis of the magnetic lensing effects to the
case of random magnetic fields, and consider also the possible
interplay between the random and uniform components. 

One may
distinguish between four different energy regimes when CRs traverse a
distance $L$ through a random field with coherence length $L_c$. If
$\delta\propto 1/E$ is the typical deflection suffered by the CRs, in
the high energy limit, corresponding to $\delta\ll L_c/L$, CRs
propagate almost rectilinearly and those arriving to an observer
from a faraway source have all seen essentially the same magnetic
field. When $\delta\sim L_c/L$, we enter the regime in which multiple
images of the same source can appear and CRs from different images
have felt uncorrelated values of the magnetic field. We find that in
this regime large magnification effects may be observable. For $\delta>
{\rm few}L_c/L$, a regime with large quantities of secondary images
appears, having the property that magnification effects tend to be
averaged out, and essentially only a blurred image with characteristic angular
size given by $\delta$ will be observable. The fourth regime is that of
spatial diffusion of the CRs, which is established when the gyroradius
of the CRs is smaller than $L_c$. For instance, for the galactic
magnetic fields this happens below the ankle (for $E/Z<0.3$~EeV), but
may happen at much larger energies for UHECRs propagating across
intergalactic magnetic fields. The energy dependence of the
associated diffusion coefficients in this regime
 can lead to important changes in the
slope of the CR spectrum, as is known to be the case for galactic CRs
below the ankle.

We will focus our work here in the systematic study of the first three
 regimes just described.
Let us mention that focusing effects produced by
random fields were previously noticed in the numerical study of
 ref.~\cite{si99}, but were 
not analyzed in detail, while the first and third regimes were considered in
refs.~\cite{wa96} in relation to the study of the time delays associated to
bursting sources.

\section{The turbulent field}

We model the turbulent magnetic field as a Gaussian random field with
zero mean and root mean square value $B_{rms}$. This can be described
by a superposition of Fourier modes as
\begin{equation}
B_i({\bf x})=\int \frac{{\rm d}^3k}{(2\pi)^3} B_i({\bf k})
e^{i(\bf{k}\cdot {\bf x}+\phi_i({\bf k}))}~,
\label{bfourier}
\end{equation}
where the phases
$\phi_i({\bf k})$ are random. If the turbulence is 
isotropic and homogeneous, the random Fourier modes satisfy
the relation
\begin{equation}
\langle B_i({\bf k}) B_j ({\bf k'})\rangle =
{B^2(k)\over 8 \pi k^2}P_{ij}(2\pi)^6 \delta ({\bf k} + {\bf k'})~,
\label{bibj}
\end{equation}
where the projection tensor $P_{ij}=\delta_{ij}-{k_ik_j/k^2}$  
guarantees that the field is solenoidal (${\bf \nabla}\cdot {\bf B}=0$)
\cite{ac00}. We will consider the general case of a 
power spectrum 
\begin{equation}
B^2(k)=B_{rms}^2 k^{-n}{(n-1)(2\pi/L_{max})^{n-1}\over 1-(L_{min}/
L_{max})^{n-1}}~,
\label{bk}
\end{equation}
for $2\pi/L_{max}\leq k
\leq 2\pi/L_{min}$, and zero otherwise. This power spectrum is already
normalized such that $\langle|{\bf B}({\bf x})|^2\rangle= B_{rms}^2$.
The particularly interesting case of
a Kolmogorov spectrum (for which the energy
density d$E/{\rm d}k\propto k^{-5/3}$) corresponds to a spectral index 
$n=5/3$.

The correlation length of this field, $L_c$, can be defined through
\begin{equation}
\int_{-\infty}^\infty {\rm d}L\langle
{\bf B}(0) \cdot {\bf B}({\bf x}(L))\rangle\equiv L_c B_{rms}^2~,
\label{lcoh} 
\end{equation}
where the point ${\bf x}(L)$ is displaced with respect to the origin
by a distance $L$ along a fixed direction.
 The integral in the lhs of Eq.~(\ref{lcoh}) can be
computed using Eqs.~(\ref{bfourier}) and (\ref{bibj}), and leads to
\begin{equation}
\pi \int_0^\infty {{\rm d} k\over k}B^2(k)=L_c B_{rms}^2 
\label{lcoh2}
\end{equation}
which can be used to express $L_c$ in terms of $L_{min}$ and
$L_{max}$ as
\begin{equation}
L_c={1\over 2}L_{max} {n-1\over n} {1-(L_{min}/L_{max})^n\over
1-(L_{min}/L_{max})^{n-1}}
\label{lcoh3}
\end{equation}
In the case of either a very sharp  $(n\gg 1)$ or of 
a very narrow-band $(L_{min}\sim L_{max})$ spectrum, one gets
$L_c\simeq L_{max}/2$, while  
for a broad-band $(L_{max}\gg L_{min})$ Kolmogorov $(n=5/3)$
spectrum, one gets instead $L_c\simeq L_{max}/5$. 

The deflection in the velocity of a particle of charge
$Ze$
travelling a distance $L$ through a magnetic field ${\bf B}$, in the
limit of small deflections, can be approximated by
\begin{equation}
\delta=\frac{Ze}{E} \left|\int_0^L {\rm d}{\bf s}\times {\bf B} (s)\right|.
\end{equation}
 For particles moving in a
turbulent field the mean deflection
vanishes. The relevant quantity is the root mean square value
\begin{equation}
\delta_{rms}^2=\langle\delta^2\rangle=
\left(\frac{Ze}{E}\right)^2 \int_0^L {\rm d}s \int_0^L {\rm d}s'
\langle{\bf B}_\perp (s)\cdot {\bf B}_\perp (s')\rangle,
\end{equation}
where ${\bf B}_\perp (s)$ stands for the component of ${\bf B}$ in the
direction orthogonal to the trajectory. This can be computed with the help of
 Eqs.~(\ref{bfourier})--(\ref{bk}). In the limit of $L\gg L_{max}$,
corresponding to the distance travelled much larger than the maximum
turbulence scale, the result can be written as 
\begin{equation}
\delta_{rms}={\frac{1}{\sqrt{2}}} \frac{Ze}{E}
B_{rms}\sqrt{L L_c}\simeq 1.5^\circ \frac{\rm{40~EeV}}{E/Z}\frac{B_{rms}}
{5~\mu{\rm G}}
\sqrt{\frac{L}{2~\rm{kpc}}}\sqrt{\frac{L_c}{50~{\rm pc}}}~, 
\label{deltarms}
\end{equation}
where we have used Eq.~(\ref{lcoh2}) to write it in terms of the 
coherence length. We have 
chosen to express its numerical value in terms of parameters relevant
for the galactic magnetic field, which has a random component
with rms strength of a few $\mu$G and a maximum scale of turbulence
of order 100 pc.

Notice that the position at which the charged particle arrives forms an angle with 
respect to its initial velocity given by
\begin{equation}
{\bf \eta}=\frac{Ze}{E}\left|\int_0^L {\rm d}s\int_0^s {\rm d} s'{\bf
B}_\perp (s')\right|, 
\end{equation}
which has a root mean square value 
\begin{equation}
\eta_{rms}=\frac{1}{\sqrt{3}}\delta_{rms}~,
\end{equation}
meaning that after traversing a distance $L$ in  a turbulent field,
the dispersion in the directions of the velocity vector is larger than
that in the position vector as seen from the departure point by a
factor $\sqrt{3}$.

\section{The high energy regime}

As a beam of charged particles propagates through a magnetic field,
its flux is focused or defocused due to the differential 
deflections of neighboring paths, with the effect being larger at
smaller energies. Eventually, the 
focusing effects may become sufficiently strong to produce multiple images of 
single CR sources.   
Heuristically, multiple images of a single source require CRs 
to reach from the source to their destination point traveling through 
uncorrelated values of the intervening magnetic field \cite{wa96}.
Consider a very distant source, so that its flux can 
be approximated as a beam of parallel CRs when they enter the region
permeated by the magnetic field. The typical condition for the formation 
of multiple images is that two initially parallel paths separated by
a distance of order  $L_c$ may reach the same point
after traversing a distance $L$ in the magnetic field, {\it i.e.}
that $\eta(L)\simeq {L_c/2L}$, or equivalently ${\delta L}/{L_c}\simeq 
{\sqrt{3}/2}$.
Multiple images of a single CR source are thus a likely possibilty 
at energies  $E\simeq\frac{2}{\sqrt{3}}E_c$,\footnote{
If the source rather than being practically at infinity were 
instead at a distance $L$ within the spatial extension of the 
magnetic field, the typical focusing condition 
is that at the midpoint the separation between alternative
paths be of order $L_c$, {\it i.e.} $\eta(L/2)\simeq L_c/L$, 
which corresponds to  $E\simeq E_c/\sqrt{6}$. This is
$2\sqrt{2}$ times smaller than in the case of a very distant source.} 
with $E_c$ defined through
\begin{equation} 
\delta_{rms}\equiv\frac{E_c}{E}\frac{L_c}{L}~.
\label{Ec}
\end{equation}
Its numerical value reads
\begin{equation}
E_c\simeq Z~41\ {\rm EeV}\frac{B_{rms}}{5\ \mu{\rm G}}
\left (\frac{L}{\rm 2\
kpc}\right)^{3/2}\sqrt{\frac{50~{\rm pc}}{L_c}}.
\label{Ecn}
\end{equation}

Notice that the heuristic argument above applies if the
magnetic field has one dominant length scale only. 
If the turbulent field has a broad-band spectrum, this heuristic 
argument should be applied to the effect
of each individual wavelength bin. Small wavelengths can actually lead to
multiple image formation at higher energies than long ones
if the amplitudes of the 
Fourier modes of the magnetic field do not decrease too fast at
small scales, as will be discussed further below. 

A detailed study of the focusing effects of the magnetic field
can be performed following the trajectories
of neighboring paths in a beam. The amplification of the flux of a 
CR source, given by the relative change in the area of the cross section
of an initially parallel beam, 
can be written borrowing the formalism familiar from
gravitational lensing \cite{mo02} as
\begin{equation}
A=\frac{1}{|(1-\kappa)^2-\gamma^2|}~.
\label{mu}
\end{equation}
Here the convergence $\kappa$ describes the isotropic (de)focusing effect while
$\gamma$ is the shear that distorts the shape of the
cross section of the beam. 
It is useful to decompose the amplification along a particular
direction of observation as $A=A_1A_2$, with $A_1\equiv
1/(1-\kappa-\gamma)$ and $A_2\equiv
1/(1-\kappa+\gamma)$ measuring the relative stretching of the beam
along the so-called shear principal axes. 
This means that a beam with an initially circular cross section 
with diameter $L_0$ develops an elliptical cross section, with one of the 
principal axes   having its length changed by $\Delta
L_1=L_0(-\kappa-\gamma)$ and the other one by $\Delta
L_2=L_0(-\kappa+\gamma)$. As the deflections due
to the random field  have no preferred 
direction\footnote{Contrary to the gravitational
lensing case, for magnetic lensing the convergence can also be
negative, and the shear can have any sign since we are not making any
distinction between the two principal axes.}, this means that
$\langle \Delta L_1\rangle=\langle \Delta L_2\rangle=0$ and 
$\langle \Delta L_1^2\rangle=\langle \Delta L_2^2\rangle$. Since
$\kappa(\gamma)\simeq -[\Delta L_1+(-)\Delta L_2]/2L_0$, it is then
easy to see that $\langle \kappa\rangle =\langle \gamma\rangle=0$ 
and  that $\langle \kappa^2\rangle =\langle \gamma^2\rangle$.

In the limit of small deflections 
the isotropic focusing of an initially parallel beam of CRs is
given by \cite{ha99}
\begin{equation}
\kappa=\frac {1}{2}\frac{Ze}{E} \int_0^L  (L-s) {\bf\nabla}\times
{\bf B}_ \cdot {\rm d}{\bf s}~.
\label{kappa}
\end{equation} 
Using eqs.~(\ref{bfourier})--(\ref{bk}) we can calculate 
$\langle\kappa^2\rangle\equiv\kappa_{rms}^2$, with the result
\begin{equation}
\kappa_{rms}=\zeta\frac{1}{\sqrt{2}}\frac{Ze}{E} B_{rms}
\frac{L^{3/2}}{\sqrt{L_c}}=\zeta\frac{\delta_{rms}L}{L_c}=
\zeta\frac{E_c}{E}~,
\label{kapparms}
\end{equation}
where we defined the numerical coefficient $\zeta$ such that
\begin{equation}
\zeta^2\equiv\frac{\pi^2}{12}
\left(\left(\frac{L_{max}}{L_{min}}\right)^{2-n}-1\right)
\frac{(n-1)^2}{n(2-n)}
\frac{1-(L_{min}/L_{max})^n}{(1-(L_{min}/L_{max})^{n-1})^2}~.
\label{zeta}
\end{equation}
In the case of either a very sharp  or of 
a very narrow-band spectrum, one gets
$\zeta=\pi/\sqrt{12}\approx 0.9$.
For a broad-band Kolmogorov 
spectrum, one gets instead $\zeta\approx 0.8~ 
(L_{max}/L_{min})^{1/6}$. 
Notice that, contrary to the case of deflections, 
small wavelengths have a significant effect upon focusing
of charged particles for spectral indices $n< 2$.
Indeed, if $n<2$ the energy at which $\kappa_{rms}$ becomes of
order unity, and thus strong lensing effects become very
likely, increases by a factor of order $\left({L_{max}}/
{L_{min}}\right)^{1-n/2}$ compared to the case of a very sharp 
spectrum.

In the high energy regime, i.e. for $E>\zeta~E_c$, 
 $\kappa$ and $\gamma$ are small and they 
can be approximated by uncorrelated random Gaussian variables with zero
mean (see Eq.~(\ref{kappa}) and take into account the fact that ${\bf B}$
is Gaussian) and dispersion $\kappa_{rms}$.  
The probability distribution of the amplification
of the beam in different observing directions,
assuming that the observer is located at the center of a sphere
 of radius $L$ filled with a random field of constant strength
$B_{rms}$ is then given by 
\begin{equation}
{{\rm d^2}p\over {\rm d}\kappa{\rm d}\gamma}=G_\kappa(0,\kappa_{rms})
G_\gamma(0,\kappa_{rms}),
\label{pkg}
\end{equation}
with $G_x(\bar x,\sigma)=\exp[(x-\bar
x)^2/2\sigma^2]/\sqrt{2\pi}\sigma$.

More interesting is to study the probability $p_s$ that the flux from a
distant source be seen  by the observer with a given magnification.
As long as we can neglect the presence of 
multiple images, which  is a reasonable approximation for $E>2 E_c$, 
this one is related to the probability in Eq.~(\ref{pkg})
by a factor $1/A$, since the magnification is just the ratio between the
observed solid angle to that actually subtended by the source (because
the surface brightness is conserved).
Hence, one has that in the high energy domain 
\begin{equation}
{{\rm d^2}p_s\over {\rm d}\kappa{\rm d}\gamma}\simeq 
{{\rm d^2}p\over {\rm d}\kappa{\rm d}\gamma}
|(1-\kappa)^2-\gamma^2|.
\end{equation}

One can then compute the probability that a source be magnified above
a certain threshold $A_0$ as
\begin{equation}
P_s(A>A_0)\simeq\int\limits_{-\infty}^\infty {\rm d}\gamma 
\int\limits_{1-\sqrt{A_0^{-1}+\gamma^2}}^{1+\sqrt{A_0^{-1}+\gamma^2}}{\rm
d}\kappa {{\rm d^2}p_s\over {\rm d}\kappa{\rm d}\gamma}.
\label{pseq}
\end{equation}
This gives that e.g. for $E=3E_c$ around 6\% of the source directions are
magnified by a factor larger than $A_0=2$, while only 1\% are
magnified by more than a factor of five.

\FIGURE{\epsfig{file=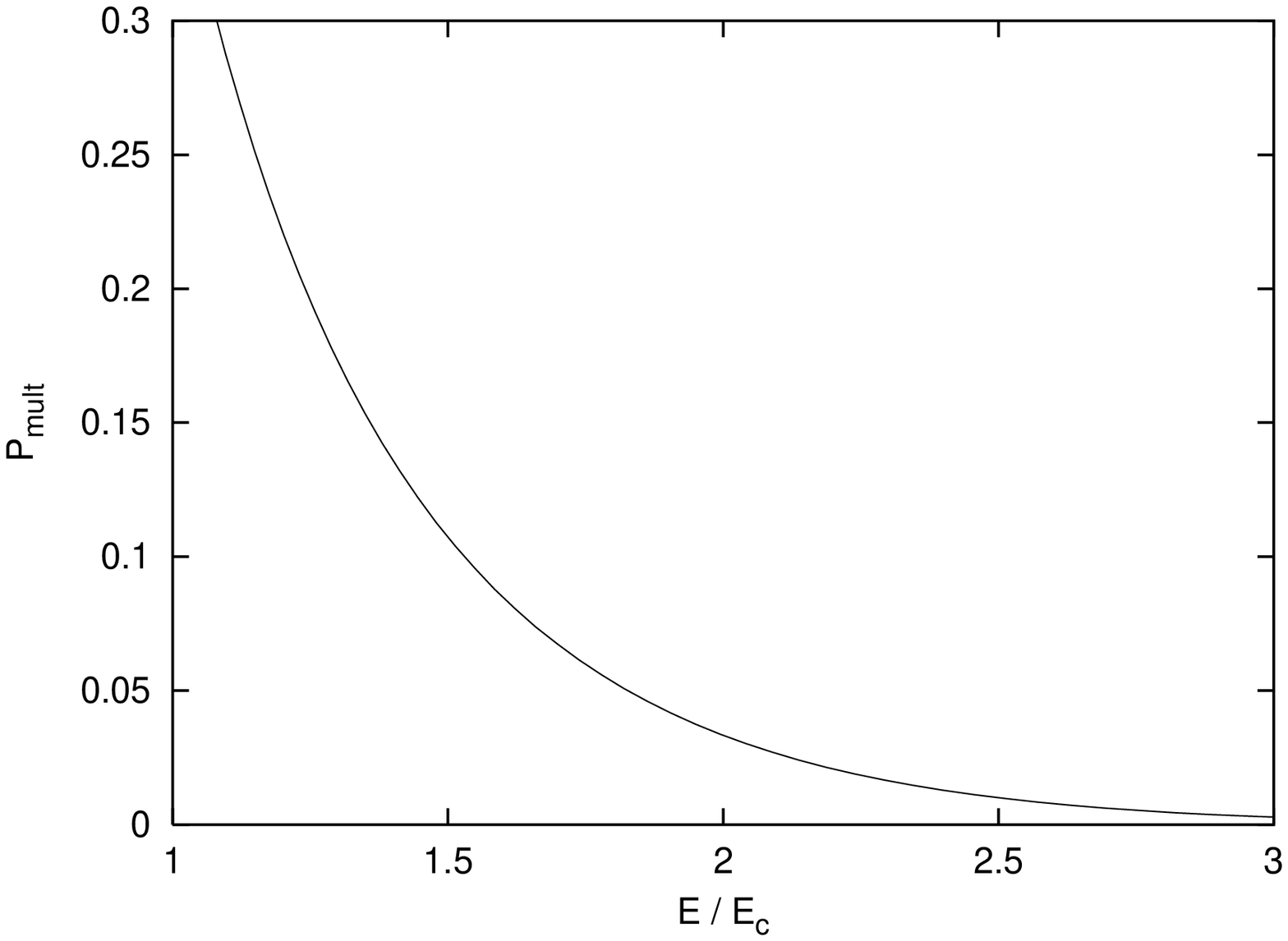,width=7truecm,angle=-90}\caption{Fraction
of the source directions for which multiple images would appear vs.
$E/E_c$.}
\label{fracmul}}

The flux from a CR source diverges after traversing a 
distance $L$ in the magnetic field for those directions such
that $(1-\kappa-\gamma)(1-\kappa+\gamma)=0$. In that case, the
source is located on top of a caustic of the magnetic field
configuration relative to the specific point of observation. 
Multiple images of a given source are visible at energies below that at 
which a caustic crosses the source location. 
The additional images exist in pairs of opposite parity, and
for the inverted image one has that $(1-\kappa)^2-\gamma^2$ is
negative. We can then compute the probability that a given source has
multiple images as
\begin{equation}
P_{mult}=P_s(|1-\kappa|<|\gamma|)=2\int\limits_0^\infty {\rm d}\gamma
\int\limits_{1-\gamma}^{1+\gamma}{\rm
d}\kappa {{\rm d^2}p_s\over {\rm d}\kappa{\rm d}\gamma}.
\end{equation}
 It has to be
noticed that this expression is reasonably accurate as long
as the fraction of sources with more than three images is negligible
(which corresponds to approximately $P_{mult}<25\%$). 
The result is plotted in Figure~\ref{fracmul}, and we see that for
$E\simeq 2E_c$ around 3\% of the sources will have multiple images and this
fraction rises to $\sim~20\%$ at $E\simeq 1.25 E_c$.

\FIGURE{\epsfig{file=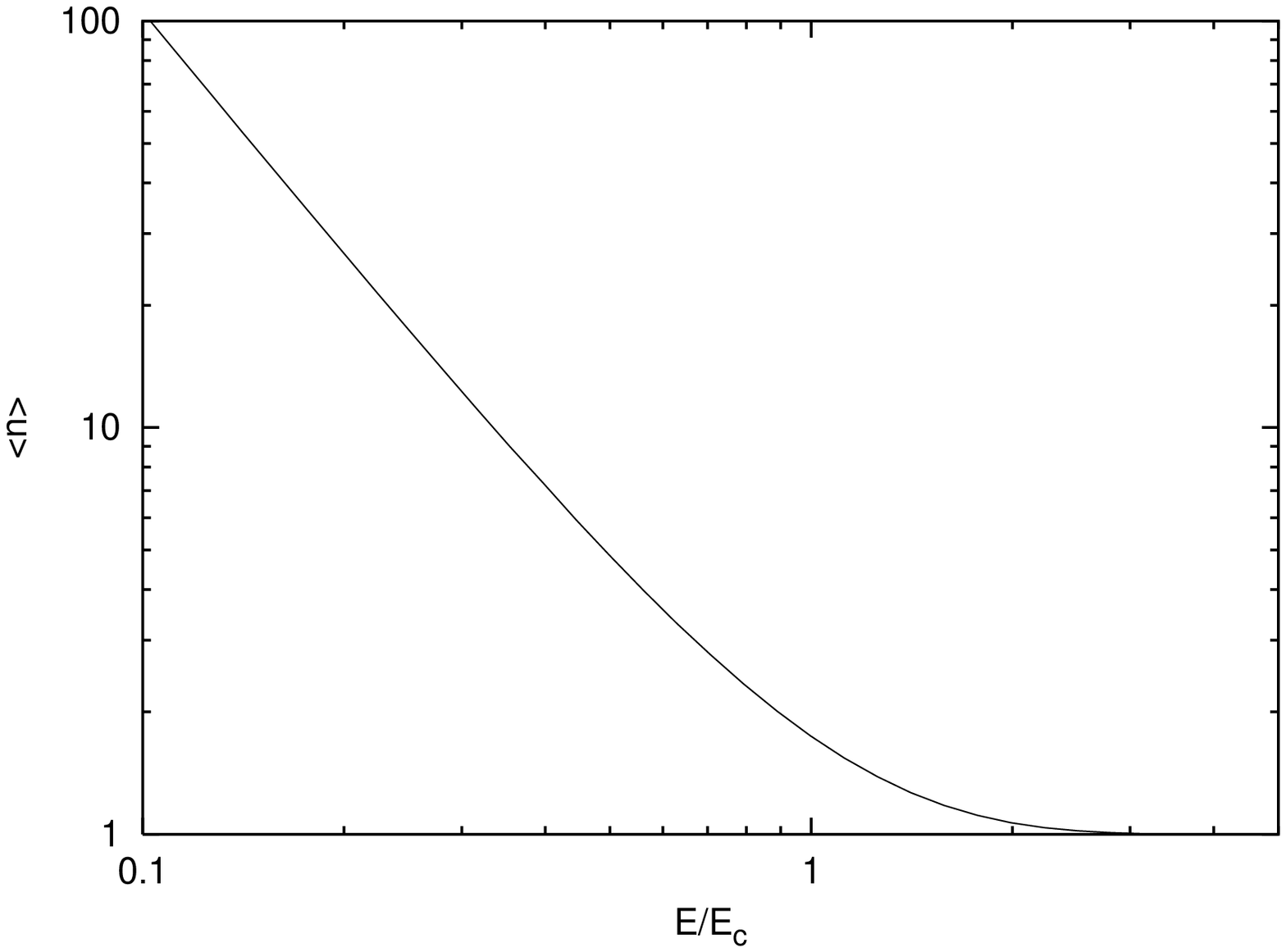,width=7truecm,angle=-90}
\caption{Mean number of images of a source vs. $E/E_c$.}
\label{nvsedec}}

For decreasing energies, the number of images associated to a source
increases. As will be shown later on in Section~\ref{scintillation}
(see Eq.~(\ref{nmedeq})), this number can be computed as the mean of
$A^{-1}$  over
all the observing directions $\theta$. Taking the mean
over many independent directions is actually equivalent to take the
mean over realizations of the random field, i.e. over different values
of $\kappa$ and $\gamma$. We then obtain
\begin{equation}
\langle n\rangle =\frac{1}{4\pi}\int {\rm d}^2\theta A^{-1}
=\int {\rm d}\kappa{\rm d}\gamma\frac{{\rm d}^2p}{{\rm d}\kappa{\rm d}\gamma}
|(1-\kappa)^2-\gamma^2|.
\label{nmed1}
\end{equation}

This expression is plotted in Figure~\ref{nvsedec} as a function of
$E/E_c$. We see that $\langle n\rangle$ takes values 1.06, 1.8 and 4.8
for $E/E_c=2$, 1 and 1/2, showing that multiple imaging is of
paramount importance for $E<2\ E_c$.

\section{Numerical results}
\label{numres}

We have implemented a numerical code to propagate charged particles
within a realization of a random magnetic field,
obtained as the superposition of 
$N$ independent waves, with random directions of
the vector ${\bf k}$ and direction of ${\bf B}({\bf k})$ randomly chosen
in the plane orthogonal to ${\bf k}$, so that the condition
${\bf \nabla}\cdot {\bf B}=0$ is automatically fulfilled. The modulus of
each mode $|{\bf B}({\bf k})|$ is drawn from a Gaussian distribution
with zero mean and dispersion given by Eq.~(\ref{bk}) with $n=5/3$
(Kolmogorov spectrum). The code allows to calculate not just deflections
of trajectories but also flux amplification along them, by following 
two additional nearby particles, using the formalism developed in
ref.~\cite{ha99}.   In practice, CR trajectories 
that reach the Earth are found backtracking trajectories of particles 
with opposite charge from the Earth out to a sphere of radius $L$.

The code results agree with the analytic 
expressions in Eqs. (\ref{deltarms}) and (\ref{kapparms})
for the rms deflections and amplifications in the small deflection
 limit for a wide range of parameters.  

A useful way to visualise the lensing properties of a magnetic
field is to plot, for a regular grid of CR arrival directions on Earth,
the incoming directions they had as they entered the
region permeated by the field. These `sky sheets' \cite{ha99} 
are quite smooth, regular and single-valued at energies  $E\gg \zeta
E_c$, meaning that only one image of each source is visible at high energies.
Its flux will be demagnified or magnified in proportion
to the amount of stretching or compression of the sheet in
the position that corresponds to the direction with which CRs enter
the magnetic field. 
For decreasing energies the sheet becomes eventually 
significantly folded  due to increased relative
deflections. CRs that enter the 
magnetic field from a direction where the sheet is folded are seen on
Earth as coming from all those directions in the grid that overlap in
that point of the picture. The additional images of the source are 
visible at energies below the one at which the fold (which moves as
as function of energy) crosses the source position.

We display these effects in Figure~\ref{sheetsM}, for three 
representative values of $E/Z$.  
The figure corresponds to the case of a field strength $B_{rms}=5~\mu$G, a
sharp spectrum ($L_{max}=L_{min}$) with  turbulence scale 
$L_{max}=100~{\rm pc}$ ($L_c=50~{\rm pc}$), and the distance
traveled by the CRs within the field is $L=2~{\rm kpc}$. These
values  are representative of the random component of the
galactic magnetic field, and we shall use them along the paper
to illustrate numerical results. The effects are rather universal
for different sets of parameters in terms of $E/E_c$, but the
angular scales involved also depend on $L/L_c$.
The simulations were performed with $N=300$ independent
waves to generate the random field, and the step with which 
the trajectories were numerically followed was 1/5 of 
$L_{c}$. 

\FIGURE{\epsfig{file=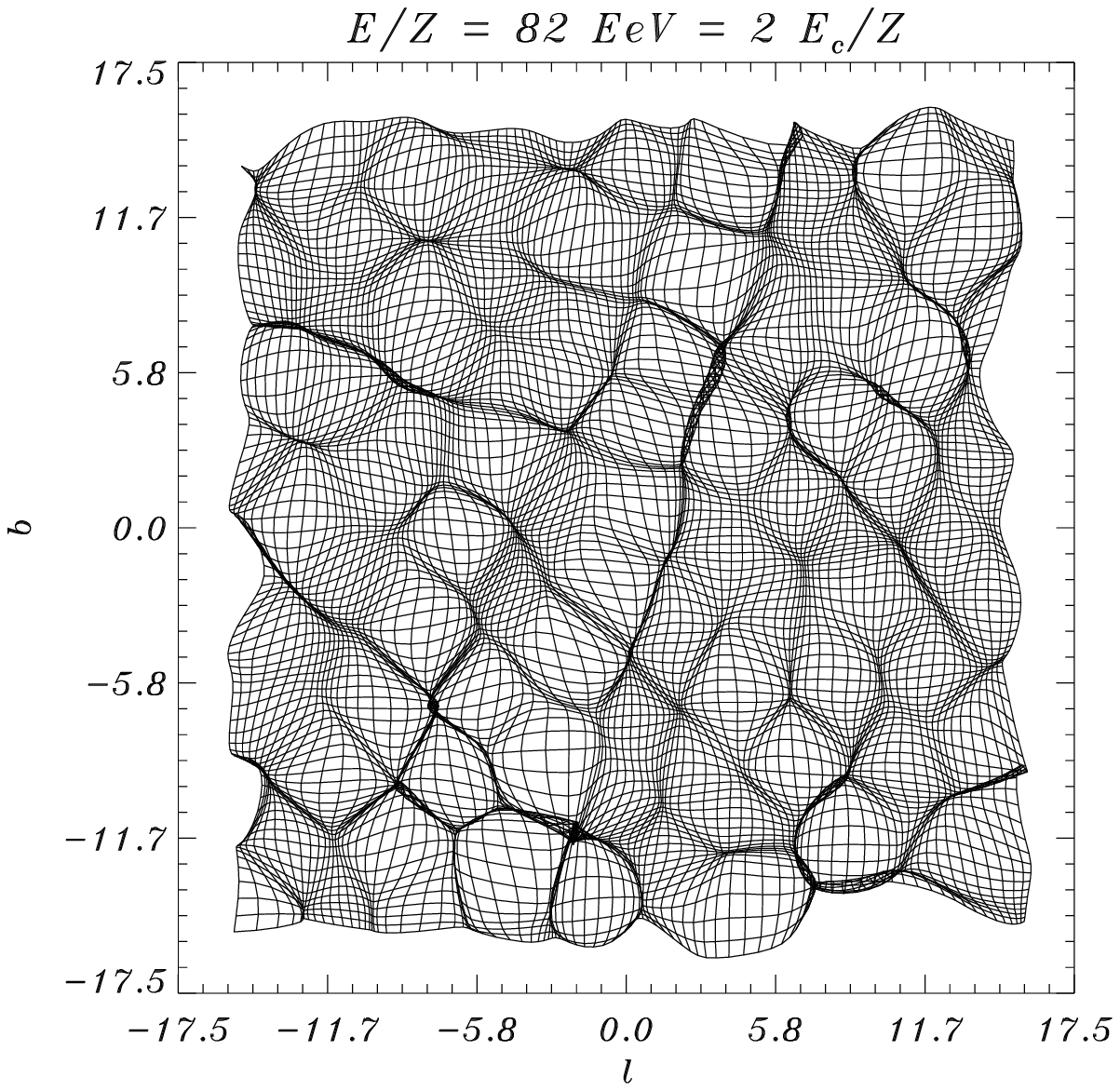,width=4.9truecm}
\epsfig{file=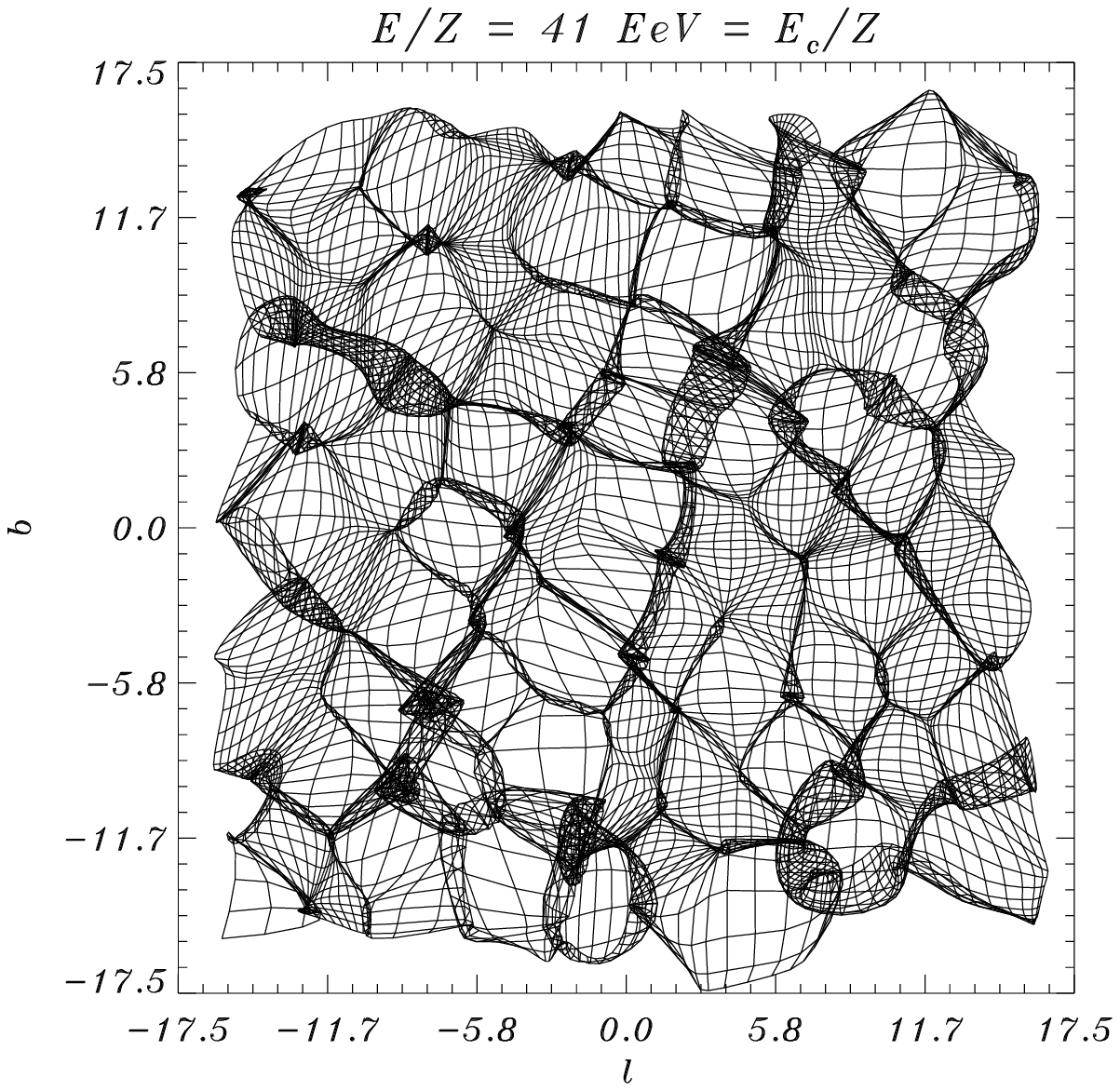,width=4.9truecm}
\epsfig{file=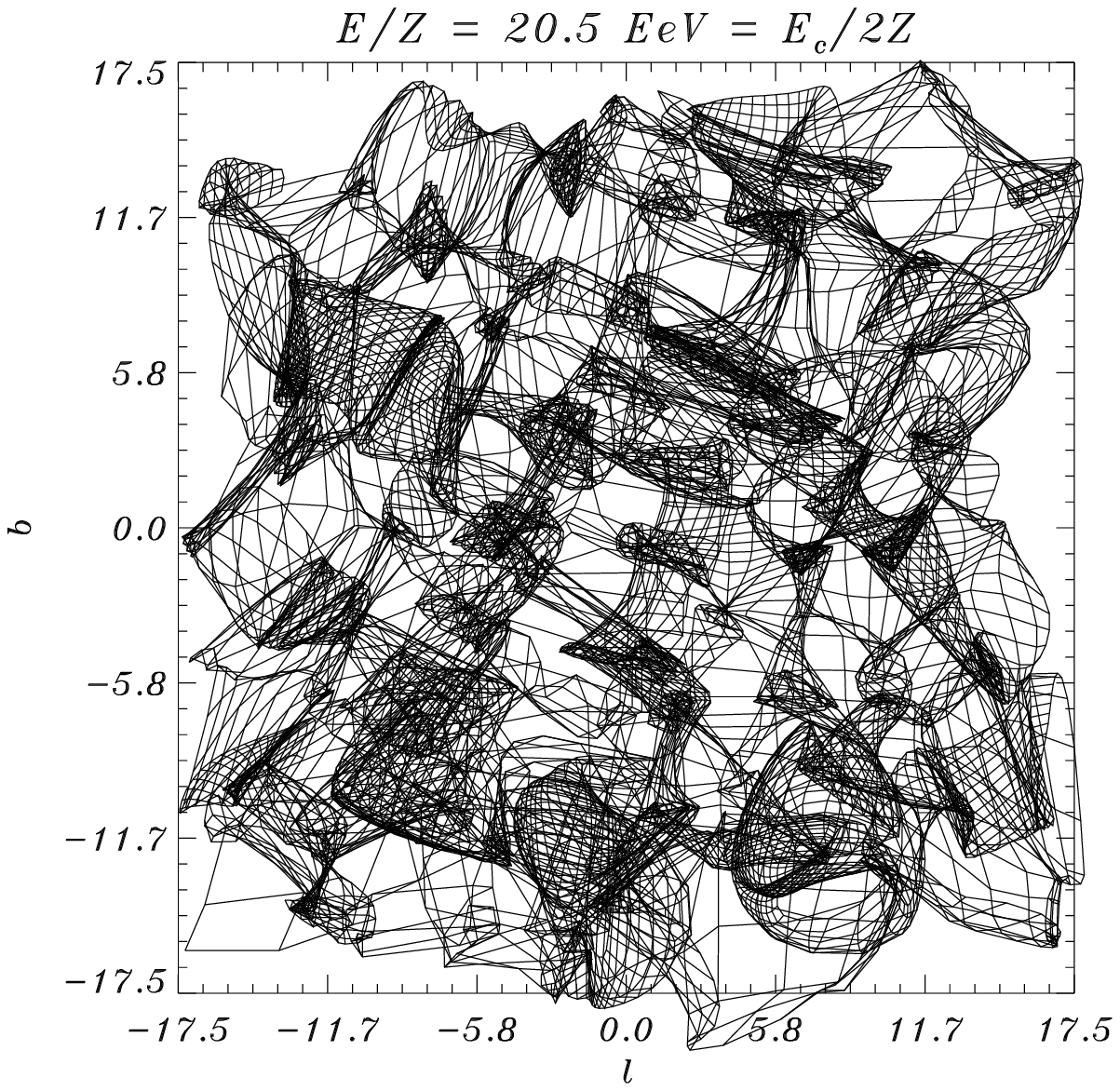,width=4.9truecm}
\caption{`Sky sheets': projection of a regular grid of CR arrival directions
on Earth onto the direction they had when they entered the region
permeated by the magnetic field. Sources located in regions where the
sheet is stretched are demagnified, and viceversa. Sources located in 
regions where the sheet is folded have multiple images. 
The parameters are $B_{rms}=5~\mu$G, $L_{max}=L_{min}=100~{\rm pc}$, 
$L=2$~kpc ($E_c=41~{\rm EeV}$).}
\label{sheetsM}}

It is clear from the left panel in Figure~\ref{sheetsM} 
that folds already cover a small but non-negligible fraction of the sky, 
of order a few \%, at energies around $2 E_c$. This fraction 
grows considerably at energies around $E_c$ (center), in agreement
with the values obtained for $P_{mult}$ in the previous Section.
At energies around $E_c/2$ (right panel) folds on top of folds 
have already developed, 
which imply sources with more than a pair of additional images.
The folds cover about half the sky at energies around $E_c/2$,
and almost the full sky at energies between $E_c/3$ 
and $E_c/4$, meaning that whatever source one choses will have already
developed multiple
images at energies larger than $E_c/4$. 

In Figure~\ref{sheetsK} we illustrate the effects of small wavelengths
in the turbulent random field through the analogous sky sheets for the case
in which the rms strength of the field is the same as above 
($B_{rms}=5~\mu$G), the distance traversed is also $L=2~{\rm kpc}$, 
but the turbulence scales extend from $L_{max}=100~{\rm pc}$ down to
$L_{min}=L_{max}/10$ with a Kolmogorov spectrum. 
In this case $L_c\approx 25~{\rm pc}$ and $E_c\approx 
58~{\rm EeV}$. We display the results for energies twice as large as
in Figure~\ref{sheetsM}, corresponding to an enhancement above $E_c$
by a factor $(L_{max}/L_{min})^{1/6}$. We used in this case 
 $N=3000$ independent waves 
to realize numerically a sufficiently random field.

\FIGURE{\epsfig{file=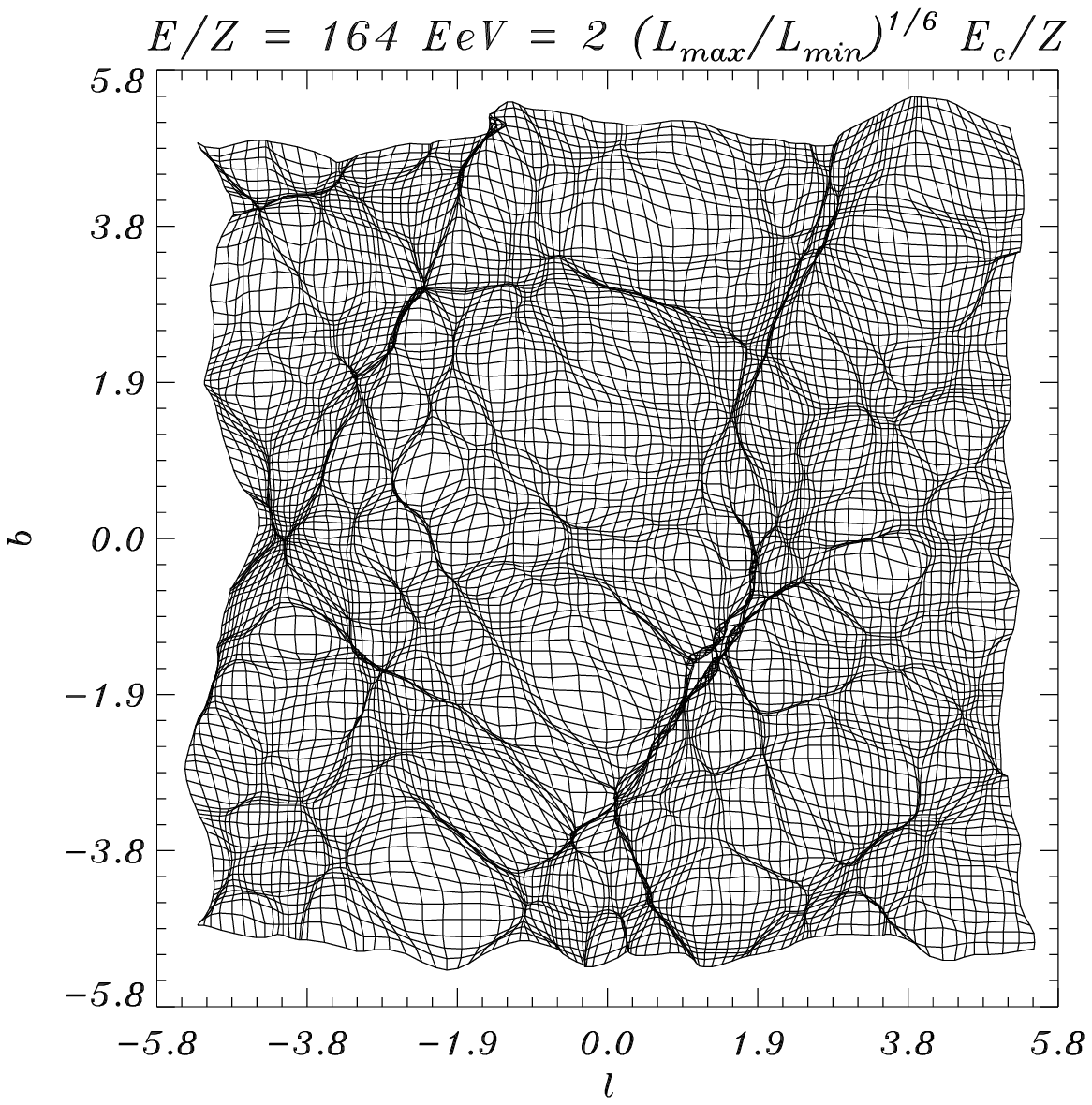,width=7truecm}
\epsfig{file=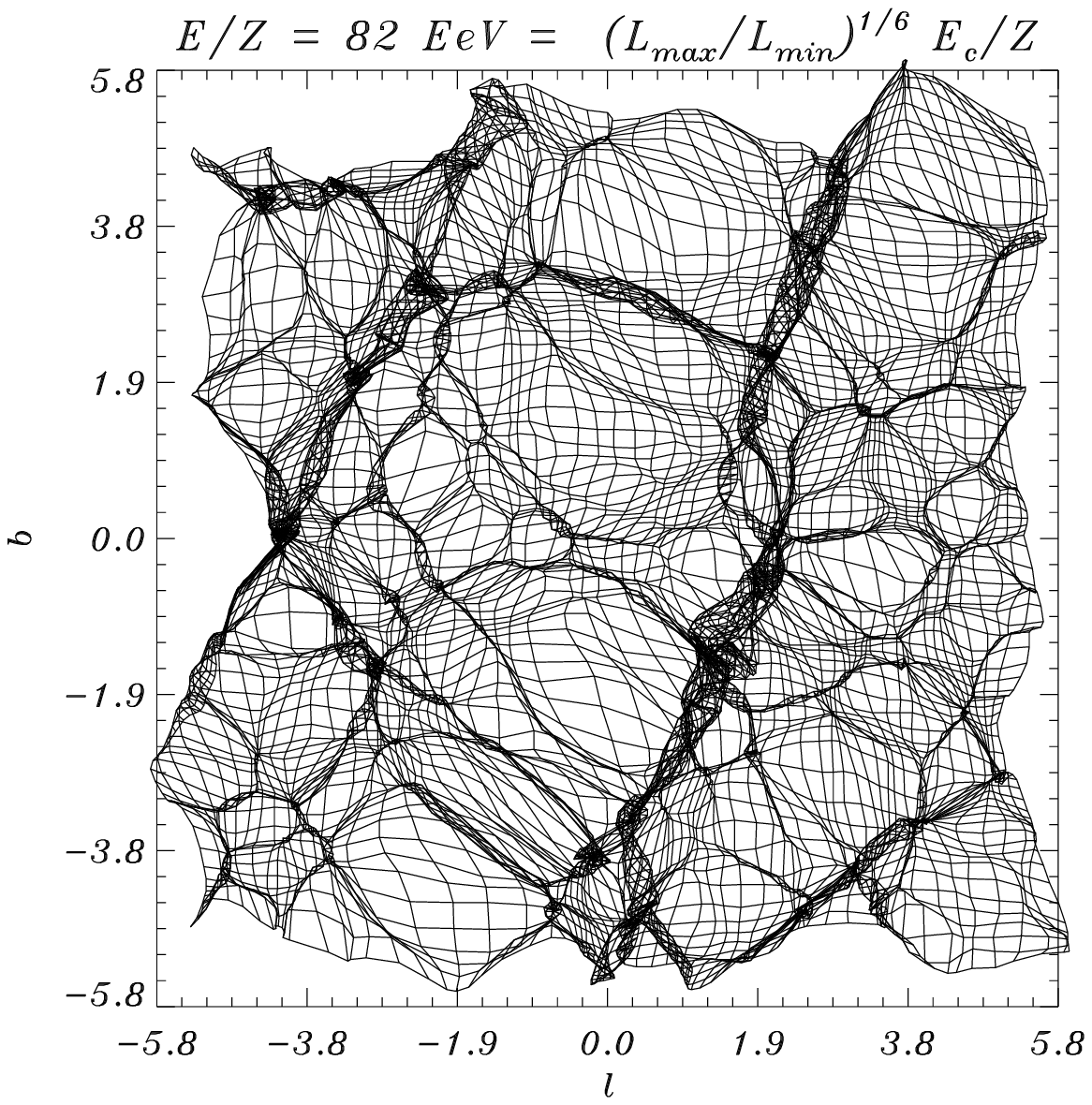,width=7truecm}
\caption{Sky sheets with the same parameters for the magnetic field 
as in Figure~\ref{sheetsM} but with a Kolmogorov spectrum for the 
turbulence, extending from $L_{max}=100~{\rm pc}$ down to 
$L_{min}=L_{max}/10$. The energies are two times larger than in 
the left and center panels in Figure~\ref{sheetsM}. Notice that the
 scales are also different, to better visualize the existence of 
structure on smaller angular scales.}
\label{sheetsK}}

The comparison between Figures~\ref{sheetsM} and \ref{sheetsK}
makes evident that the short wavelength modes of the
turbulence lead to folds at larger energies than the long
wavelength modes, but their angular scale is smaller (notice that the
linear scales in Figure~\ref{sheetsK} are three times larger than
in Figure~\ref{sheetsM}). However, as we shall discuss later, the integrated
magnification effects upon a source depend not only on its proximity
to a fold but also on how fast the fold moves across the source location,
which is basically determined by the long wavelength modes, and this may
average out the effects associated to the  short wavelengths.

The numerical code can also follow the change in apparent position and the
flux amplification of a given source image as a function of energy,
exploring at each energy step a neighborhood of arrival directions
from the previous step, and backtracking it until it reaches the
appropiate direction to the source with the required accuracy. 
The arrival directions of multiple images at some fixed energies
found in the data used to plot the sheets above are used as starting
points to follow them in energy. In Figure~\ref{spectra} we display
the amplification as a function of energy for two illustrative 
diverse behaviours. 

The solid line in Figure~\ref{spectra} is illustrative of the effects 
that sources
located in only a small fraction of the sky, of order 2\% or less,
undergo due to magnetic lensing. This source is close to
the location where a fold in the lens mapping develops at relatively
high energies. Its principal image is highly magnified over a
significantly broad energy range. Multiple images are visible
at relatively high energies, with magnification peaks that have 
strong integrated effects. Indeeed, the magnification of the secondary
images behaves around the energy $E_f$ at which the fold crosses the
source location as $A=A_E/\sqrt{1-E/E_f}$ \cite{ha00}, with a coefficient $A_E$
that we found to be typically of order but larger than unity if $E_f$ is 
above $2~E_c$. Since the magnification integrated over an energy 
bin of order 10\% of the peak location is around $12~A_E$ (see next
Section),  we conclude
that quite strong features are certainly imprinted upon the observed
spectra of these sources, albeit with a small probability.
The principal and the first pair of secondary images are demagnified at
decreasing energies, but many additional images (not depicted here) 
will eventually become
visible, although with much narrower peaks.  

\FIGURE{\epsfig{file=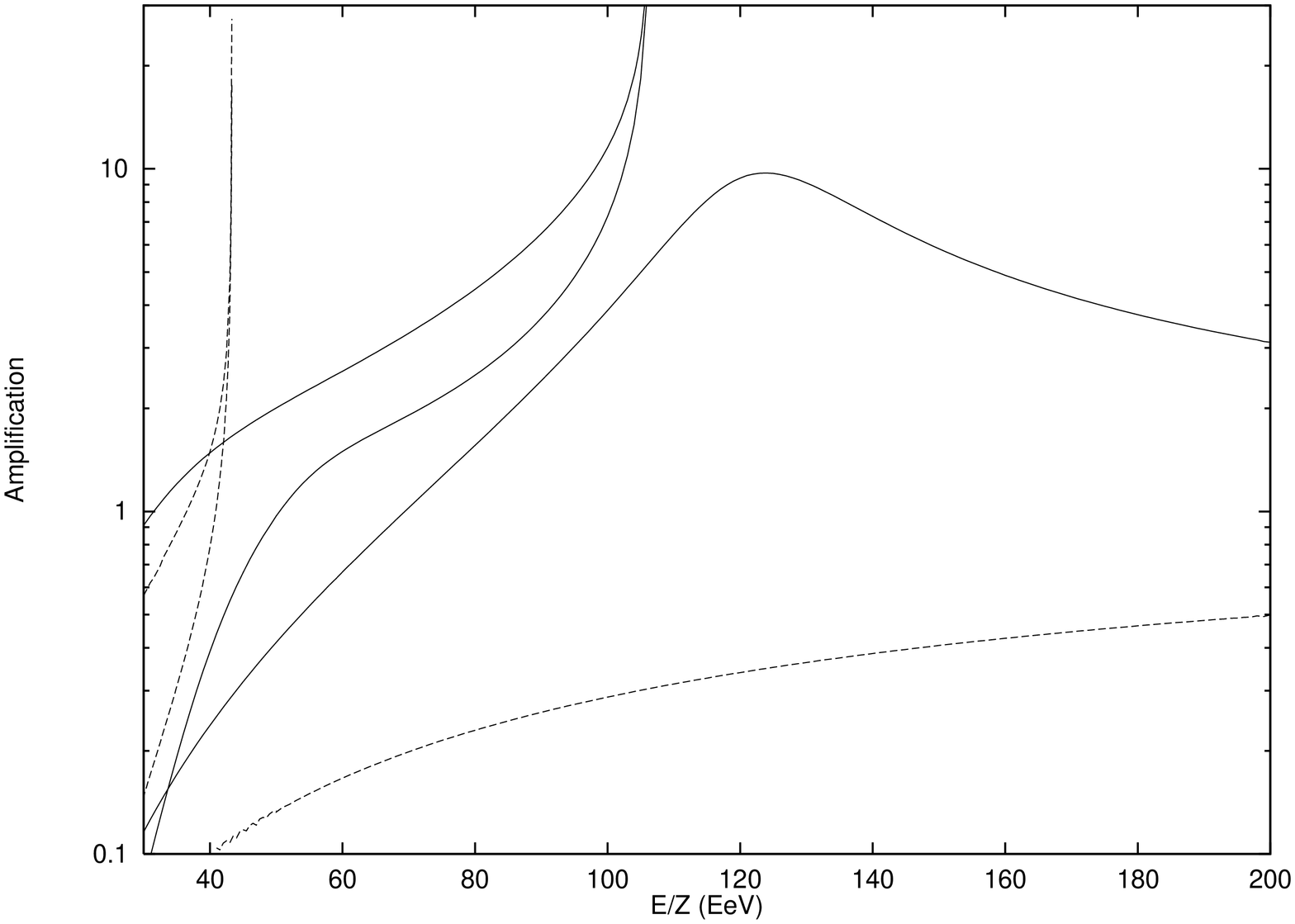,width=7truecm,angle=-90}
\caption{Amplification vs. energy for the principal and first
pair of multiple images of a source location close to where a
fold develops at energy above $2~E_c$ (solid lines) and for a 
source location which is first crossed by a fold at energy
around $E_c$ (dashed lines). The parameters are the same as 
in Figure~\ref{sheetsM} ($E_c=41~{\rm EeV}$).}
\label{spectra}}

The dashed lines in Figure~\ref{spectra} are illustrative of 
a more likely situation. The first pair of secondary images of this 
source develops at an energy around $E_c$. At 
relatively high energies the source location is away from a fold,
in a stretched region of the sky sheet, and thus its principal 
image is demagnified. The magnification peaks of the first
pair of secondary images are relatively narrow, but still lead to 
significant integrated effects, since typically $A_E$ is around 0.5 
if $E_f$ is around $E_c$. We find that typical values of $A_E$ 
remain above 0.1 (below which no significant effects would be observable unless
the experimental energy resolution were better than 10\%) 
down to energies $E_f$ around $E_c/3$. 

At energies below $E_c/3$ it becomes very likely that sources
develop a high number ($\langle n\rangle>10$) 
of secondary images. It is thus unpractical
to trace each individual image apparent location and amplification
as a function of energy. A ray shooting technique becomes instead
more appropriate. In such simulations, a large number of antiparticles 
are thrown isotropically from Earth, and those that after traversing 
a distance $L$ point to a direction closer than a required accuracy
from that to a fixed `source' are recorded. The ratio of this number
to the one that would have been obtained in the absence of magnetic
deflections is just the corresponding magnification of the total 
source flux, summed over all images. For a fixed number of rays
shooted, a smaller source size degrades the precision with which
demagnifications are recorded. A large source size instead 
prevents recovering large magnification peaks, due to averaging
effects.

\FIGURE{\epsfig{file=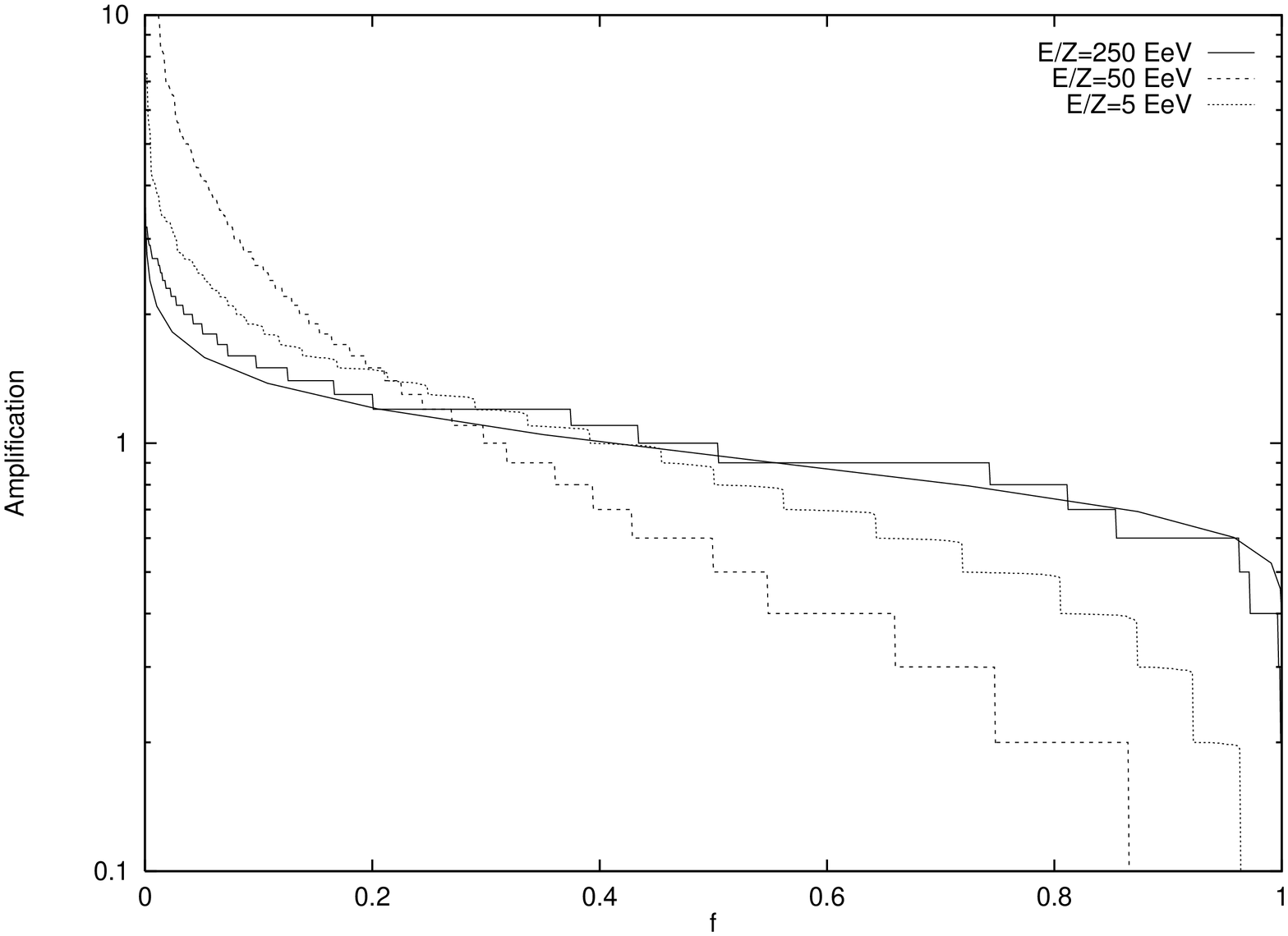,width=7truecm,angle=-90}
\caption{Fraction of sources (in the horizontal axis) with amplification
higher than a given threshold (in the vertical axis), obtained from the
ray-shooting simulation described in the text, for representative
values of $E/Z$. The fit to Eq.~(\ref{pseq}) is also shown for the highest
energy. The magnetic field parameters are as in Figure~\ref{sheetsM}.}
\label{fraccion}}

In Figure~\ref{fraccion} we plot the result of evaluating through
such ray shooting technique the amplification of 1080 consecutive
source locations of angular diameter 1/3 of a degree each. The parameters
are the same as in previous Figures. We plot the magnification
of each image sorted from higher to lower values, with the horizontal axis
rescaled to 1. Thus, the value of the abscisa for a given amplification
$A_0$ corresponds to the fraction of sources that have amplification 
larger than $A_0$. The results are displayed for three representative
values of $E/Z$. At energies $E\gg E_c$ about half the sources are 
sligthly magnified while the other half are slightly demagnified, in
comparable proportions. The probabilities here are in agreement with
Eq.~(\ref{pseq}), which is also plotted for comparison.
 At decreasing energies more sources are 
demagnified rather than magnified, and there is a significant enhancement
of the large magnification tail. The large magnification tail reaches
its maximum strength around $E\approx Z~ 50~{\rm EeV}\approx E_c$.
The fraction of sources demagnified also reaches its maximum, of order
2/3, at $E\approx E_c$. At energies below $E_c$ the curve starts to
level off again, as we enter some kind of scintillation regime where 
each source has a very large number of demagnified images with total
amplification of order unity. Only the very few sources that have images 
with extremely narrow peaks precisely at the energy under consideration
make a small contribution to the high magnification tail (but which is
here somewhat suppressed by finite source size effects). 

In the next two Sections we show how these issues can be addressed 
analytically.

\section{The appearance of new images}

As we have seen from the previous simulations, new images of CR
sources can appear when the mapping from the source's plane to the
observer's plane becomes multiply valued or, in other words, when the
observer's sky becomes folded when projected into the source's sky
(Figures~\ref{sheetsM} and \ref{sheetsK}).  This pictorial view 
of a folded sky is quite
useful to study the general properties of the images in the regime of
multiple imaging. To see this let us consider 
the deflection of CR trajectories after traveling a distance $L$
through a turbulent magnetic field and let us analyse the properties
of the mapping
between the incident directions ($\beta$) and the observed ones
($\theta$). At large energies, trajectories are straight lines, and
the mapping is the identity, but the typical deflections increase with
decreasing energies and make this mapping non-trivial. 
Since directions separated by more than $L_c/L$ probe
uncorrelated values of the magnetic field, this means that they suffer
uncorrelated deflections. These uncorrelated deflections have
a random distribution of directions and a typical amplitude given by 
$\delta_{rms}$. The two-dimensional deflection process can
be thought as the superposition of one-dimensional deflections in two
orthogonal directions, and we show in Figure~\ref{mapping} a picture
of a plausible mapping for one of the 
directions. 

\FIGURE{\epsfig{file=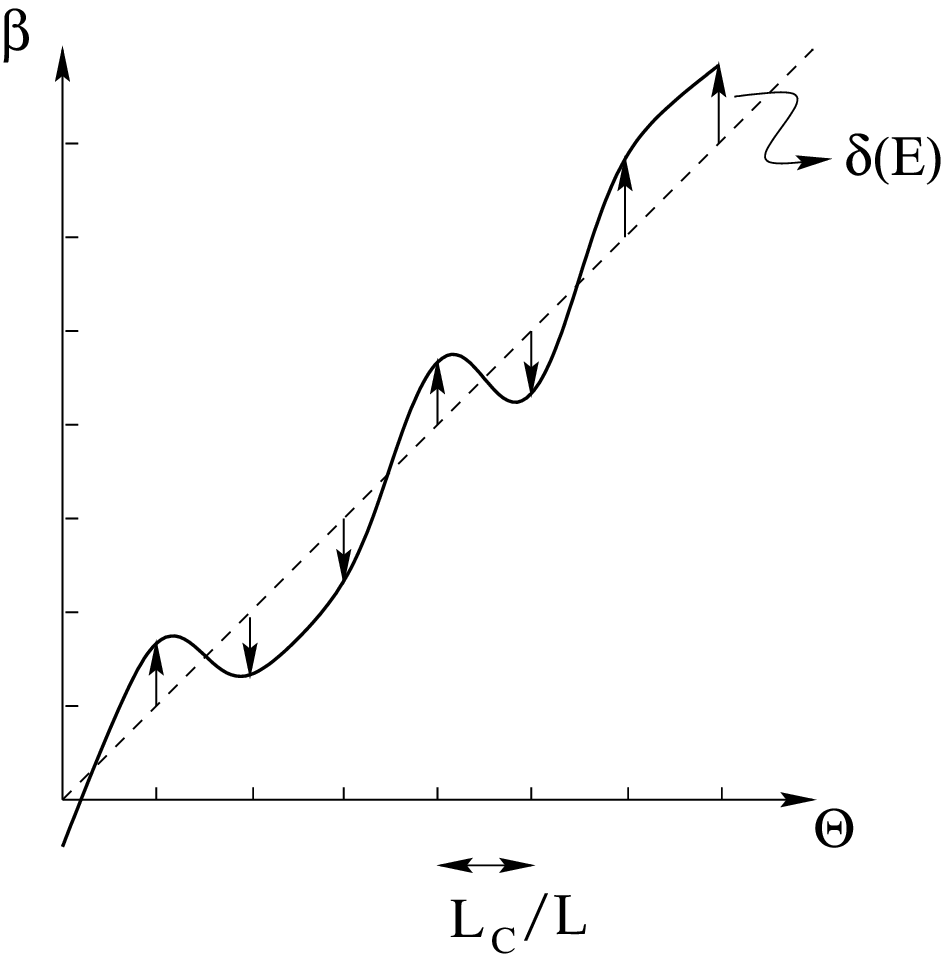,width=9truecm}
\caption{Pictorial view of the mapping between the arrival direction
of cosmic rays at the observer ($\theta$) and the incident direction
from the source ($\beta$).}
\label{mapping}}

The general properties of this mapping can be simply
understood  considering a network of directions separated by
$L_c/L$ and assuming that these points are deflected with a certain
(energy dependent) amplitude in either direction. Two neighboring
points  in this network
can then be deflected towards the same or in opposite
directions. If they are deflected towards the same direction, essentially  only the
position of the image changes. On the other hand,
if two neighboring directions are deflected toward each
other (so that $\Delta\beta<\Delta\theta$), 
a caustic will eventually form in between them (when $\partial
\beta/\partial\theta=0$) as the energy
decreases, leading to large amplifications of the flux and to the
formation of multiple images. Since in this picture 
neighboring directions have
 a probability $1/4$ of approaching each other, the
typical separation between the directions where caustics will form is
$\Delta\beta= 4 L_c/L$, in agreement with the results of the detailed
simulations.  When neighboring points are deflected in opposite
diverging directions (so that $\Delta\beta>\Delta\theta$), 
a low magnification region results. This low magnification regions
will appear always in
between the caustics.

Around a direction where a caustic forms, which is the most
interesting case due to the possibility of having large
magnification effects, the mapping in the
direction orthogonal to the fold can be approximately
described by a cubic polynomial. Defining $y\equiv(\beta-\beta_0) 
L/L_c$ and $x\equiv(\theta-\theta_0) L/L_c$ (where 
$\beta_0$ is the direction where the caustic forms and  $\theta_0$ 
its image, i.e. the corresponding critical line in  
the observer's plane), the mapping  can then 
be written as\footnote{For
simplicity we omit a quadratic term, what corresponds to the
assumption that the positive and negative deflections generating the
folds have similar magnitudes. The generalisation to the asymmetric
case can easily be done by assuming that the symmetry point of the
fold in the source plane, $\beta_0$, moves with energy. The leading
correction would then be obtained replacing
$\beta_0\to\beta_0+D(E-E_\ast)$, with $D$ a constant. The main effect
of this change is to modify the energy gap between the first bump
in the spectrum and the appearance of the first fold peaks, as well as the
energy width of these last (see below).}
\begin{equation}
y=ax+bx^3.
\label{cubic} 
\end{equation}

\FIGURE{\epsfig{file=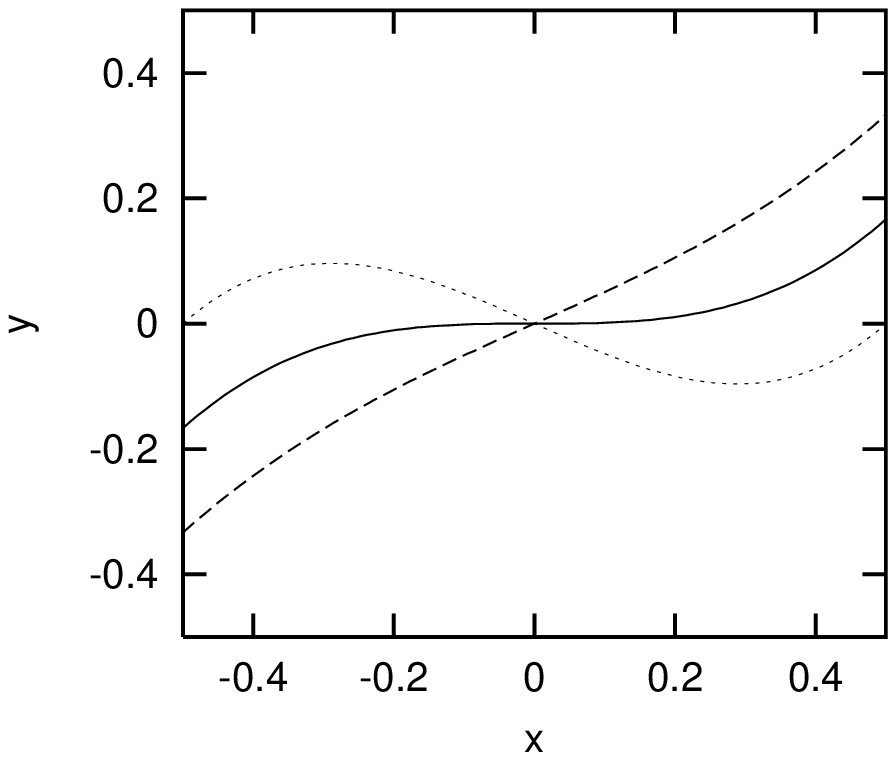,width=9truecm}
\caption{Mapping near a caustic for three different values of
$E_\ast/E=0.5$ (dashed), 1 (solid) and 1.5 (dotted).}
\label{caust}}

The coefficients $a$ and $b$ are functions of the deflections $\delta(E)$.
For $E\rightarrow\infty$, we have $a\rightarrow 1$ and
$b\rightarrow 0$, and it is natural then to take that 
$a-1 \propto b \propto \delta
\propto E^{-1}$. The caustic forms at
$x=0$ when $\partial y/\partial x=0$. 
This means that $a$ can be written as $a=1-E_\ast/E$, where $E_\ast$
denotes the energy at which the caustic forms, and $b=CE_\ast/E$ with
$C$ being a constant.
The value of $C$ can be inferred from the fact that the slope 
$\partial y/\partial x$ has to be unity  at some point near
$x\simeq\pm1/2$. Hence, we should have 
 that typically $C\simeq 4/3$. The mapping $y(x)$ is depicted in 
Figure~\ref{caust} for different values of the energy, 
showing the formation of caustics as the energy decreases below $E_\ast$. 
Notice that the energy $E_\ast$ will be different for different source
locations, since it is determined by the energy of formation of the fold
nearest to the source chosen. An average value of $E_\ast$ can be
estimated from the fact that $y(x=1/2)=1/2-E_\ast/3E\simeq 1/2-\delta
L_c/L$, and as the rms value of the one dimensional deflection
$\delta$ is just $\delta_{rms}/\sqrt{2}$ this leads to $\langle
E_\ast\rangle\simeq (3/\sqrt{2})E_c$.

For $E<E_\ast$, a couple of folds appear that move appart as the
energy decreases. 
Their location  in the source plane, $\pm y_f$, can
easily be found from the condition $\partial y/\partial x|_{y_f}=0$,
leading to
\begin{equation}
y_f=\frac{1}{3}\frac{E_\ast}{E}\left(1-\frac{E}{E_\ast}\right)^{3/2}.
\label{yfold}
\end{equation}

Particles from an incident direction $y_s$ will be observed from the
direction(s) $x$ that solve the cubic equation (\ref{cubic}).
 The solution is unique (just one image of the source) for $E>E_f$,
where $E_f$ (which is smaller than $E_\ast$) is the energy
for which the fold location coincides with the source direction. This
means that $E_f$ is just obtained from the relation
\begin{equation}
y_s=\frac{1}{3}\frac{E_\ast}{E_f}\left(1-\frac{E_f}{E_\ast}\right)^{3/2}.
\end{equation}
 The location of the image in this case is given by 
\begin{equation}
x=\left(-q+\sqrt{Q}\right)^{1/3}+\left(-q-\sqrt{Q}\right)^{1/3}, 
\label{onesol}
\end{equation}
where we introduced $q\equiv -y/(2b)$, $p\equiv a/(3b)$ and $Q\equiv
p^3+q^2$.

The amplification of an image (in the direction orthogonal to the
caustic) is given by 
\begin{equation}
A_\perp=\frac{\partial\theta}{\partial\beta}
=\left(\frac{\partial y}{\partial x}\right)^{-1}=(a+3bx^2)^{-1}
\end{equation}
Using Eq.~(\ref{onesol}) we then get 
\begin{equation}
A_\perp=\frac{1}{6b\sqrt{Q}}\left[
\left(-q+\sqrt{Q}\right)^{1/3}-\left(-q-\sqrt{Q}\right)^{1/3}\right].
\label{aperp1}
\end{equation}
Notice that in the demagnification regions, the deflections will just have
the opposite sign than what was assumed before, 
and hence a similar expression will hold for the
amplification but changing $E\rightarrow
-E$ in the expressions for $a$ and $b$. 

When the source is located within the folded region, i.e. for
$|y_s|<y_f$, there are three different images, whose positions can
be written as
\begin{equation}
x^{(k)}=\sqrt{-\frac{4a}{3b}}\cos\left[\frac{1}{3} \arccos\left(-\frac{3y_s}{a}
\sqrt{-\frac{3b}{4a}}\right)+\frac{2k\pi}{3}\right],
\end{equation}
with $k=0,1,2$. Here $k=0$ corresponds to the principal image (the one
already present at high energies), while $k=1,2$ are the pair of
images created when the fold crosses the source location (i.e. for $E<E_f$).
The amplification of these images is
\begin{equation}
A^{(k)}_\perp=\left(1-\frac{E_\ast}{E}\right)^{-1}
\left(1-4\cos^2\left[\frac{1}{3}
\arccos\left(\frac{3y_sE/E_\ast}{(1-E/E_\ast)^{3/2}}\right)+
\frac{2k\pi}{3}\right]\right)^{-1}
\label{ampE}
\end{equation}
In general the total amplification will be  the product of 
the amplification in the
direction transverse to the fold, $A_\perp$, times the amplification 
in the direction along the fold, $A_\parallel$ (that can be taken as
constant  in first
approximation).  Then, one has that $A_{tot}=A_\parallel \sum_k
|A^{(k)}_\perp|$. We notice that a negative value of $A^{(k)}_\perp$
would just mean that the corresponding image is inverted, i.e. that it
has negative parity.

\FIGURE{\epsfig{file=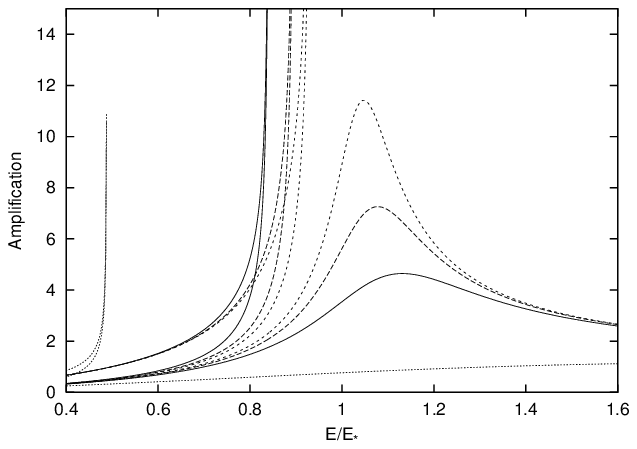,width=9truecm}
\caption{Amplification of a point source for different values of $y_s=
1/160$ (short dashes), $1/80$ (long dashes), $1/40$ (solid) and $1/4$ (dotted).}
\label{lomas}}

In Figure~\ref{lomas} we show the transverse magnification of the images of
a source for different values of $y_s$ (i.e. for different values of
 the distance from the source
to the symmetry point of the fold  in units of $L_c/L$). 
When plotted in terms of $E/E_\ast$, these curves
only depend on the value of $y_s$,
with  smaller  $y_s$ leading to more pronounced magnification effects.
Notice the similarity of Figure~\ref{lomas} with the energy dependence
of the amplification in the numerical examples in Figure~\ref{spectra}. 
 
Let us now first concentrate in the magnification of the principal
image of the source. We find that for small values of $y_s$ ($y_s<
0.1$), sizeable peaks are observed.  
 The theoretical amplification obtained from Eq.~(\ref{aperp1}) 
 provides an excellent fit to the amplification of these peaks obtained in 
numerical simulations,  as is exemplified in
Figure~\ref{lomafit}. Notice that the value of $A_{max}$ determines
$y_s$, while the location of the peak fixes $E_\ast$.

\FIGURE{\epsfig{file=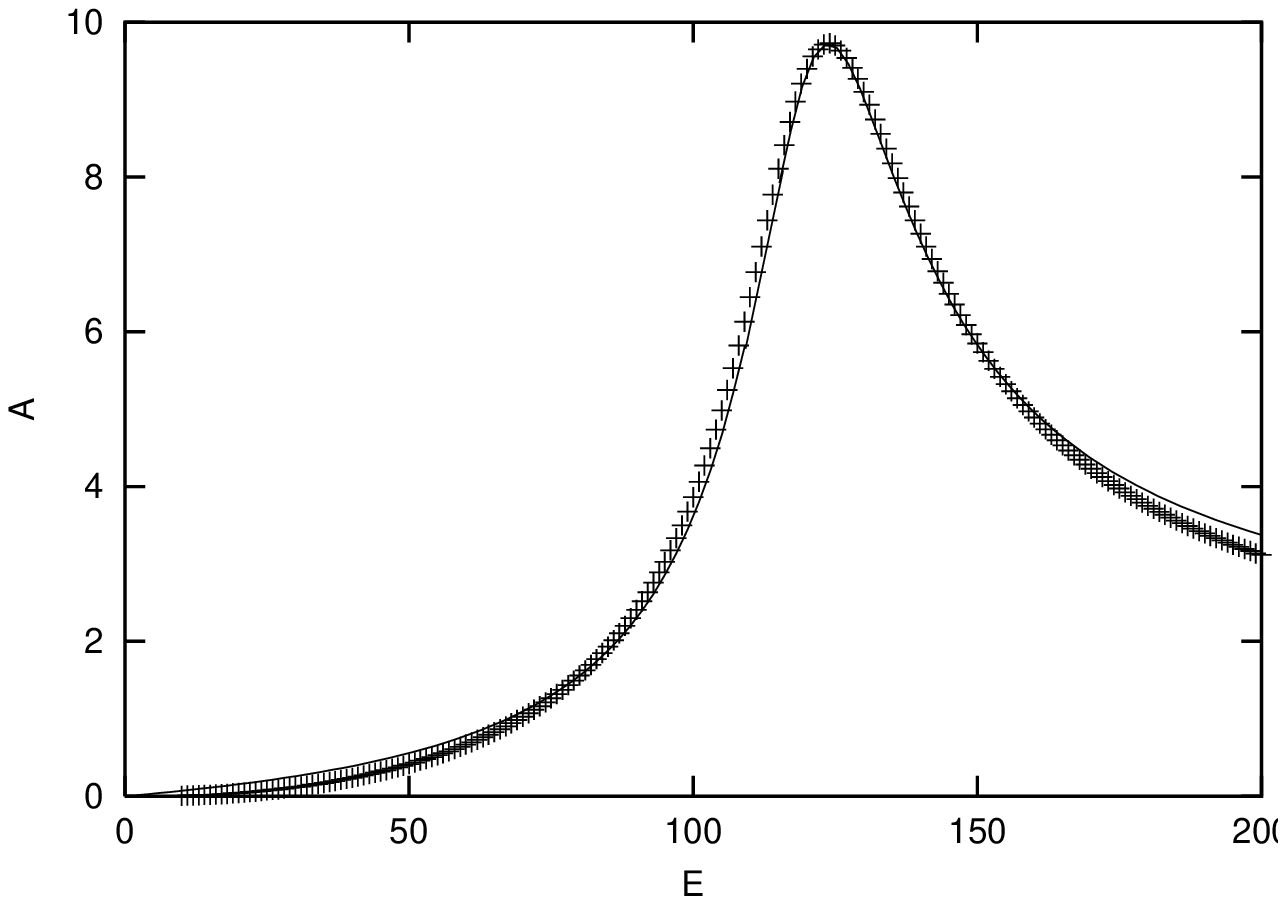,width=9truecm}
\caption{Amplification of the principal image of a point source computed
numerically (+) and fit using the theoretical model.}
\label{lomafit}}

The relation between the peak magnification achieved, $A_{max}$, and the
source location $y_s$
is depicted in  Figure~\ref{amax.fig} (solid line). This curve is
accurately  fitted, for $y_s<0.1$,  by the expression
\begin{equation}
A_{max}\simeq 2\left(\frac{0.086}{y_s}\right)^{2/3},
\label{amax}
\end{equation}
as is also illustrated in that Figure (dashed line).

The peaks in the magnification of the principal image for $y_s<0.1$
are quite wide, having typical widths of $10\div 30$\% of the peak
energy. This is very interesting because it means that these peaks
have a width comparable to the energy resolution of UHECR experiments,
and should then be in principle resolvable.

We can also estimate the probability that the peak magnification of
the principal image  be
larger than a given value $A_0$. This equals the probability that the
source be at a distance to the caustic formation line smaller than
$y_0L_c/L$, with $y_0$ related to $A_0$ through
Eq.~(\ref{amax}). Assuming that a rectangular network of caustics
is formed in the sky  separated among them  by $4L_c/L$
(i.e. by $\Delta y\simeq 4$), this
probability will just be 
equal to the fraction of the sky covered by a network of strips
of width $2y_0L_c/L$, and separated among them  by $4L_c/L$, which is 
just equal to $y_0$, i.e.
\begin{equation}
P(A_{max}>A_0)=P(y<y_0)=y_0\simeq 0.24 A_0^{-3/2},
\end{equation}
with the last equality being valid for $A_0>2$.
This expression was checked numerically 
by following the principal images of a thousand
randomly located sources and computing their amplification, and it
indeed agrees very well with the numerical results obtained.
These results imply that there is for instance a one percent
probability for the principal image to be magnified by more than a
factor eight, while the probability for it to be magnified by more than a
factor of three is $\sim 5$\%.

\FIGURE{\epsfig{file=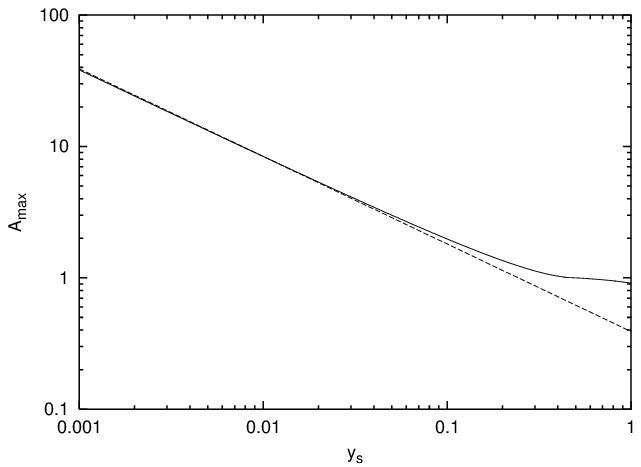,width=9truecm}
\caption{Maximum amplification of the principal image of 
a point source as a function of $y_s$ (solid line). 
The dashed line corresponds to
the fit described in the text.}
\label{amax.fig}}

Turning now to discuss the magnification of the new pair of images
appearing below $E_f$, we see from Figure~\ref{lomas} that they have
associated divergent peaks, which for increasing values of $y_s$
appear at smaller values of $E_f/E_\ast$ and become increasingly
narrower. We can obtain an analytic approximate expression for the
amplification of these peaks making a Taylor expansion of Eq.~(\ref{yfold})
around the fold location $x_f=\pm\sqrt{-a/3b}$. This gives (for the
fold at positive values of $y$)

\begin{equation}
y\simeq y_f+\frac{1}{2}y_f''(x-x_f)^2+\dots ,
\label{yexp2}
\end{equation}
with
\begin{equation}
y_f''\equiv\left.\frac{{\rm d}^2y}{{\rm d}x^2}\right|_{x_f}=
-4\frac{E_\ast}{E}\sqrt{1-\frac{E}{E_\ast}}.
\end{equation}
Deriving the expression Eq.~(\ref{yexp2}) 
we obtain the amplification of the flux of CRs
of energy $E$ coming from a source in the direction $y_s$, which is
\begin{equation}
A_\perp=\left(\frac{\partial y}{\partial x}\right)_{y_s}^{-1}
\simeq \pm \left( 2y_f''
(y_s-y_f)\right)^{-1/2}
\end{equation}
Furthermore, since 
the fold moves with energy according to Eq.~(\ref{yfold}) while the
source is at a fixed position $y_s$, we can write $y_f\simeq
y_s+({\rm d}y_f/{\rm d}E)_{E_f}(E-E_f)$, with
\begin{equation}
\frac{{\rm d} y_f}{{\rm d}E}=-\frac{1}{3E}\left[\frac{1}{2}+\frac{E_\ast}{E}\right]
 \sqrt{1-\frac{E}{E_\ast}}.
\end{equation}
 Combining these expressions, one gets for
energies close to $E_f$ that\footnote{Notice that in this
approximation the magnification of the two images is similar and their
difference only shows up if we include additional terms in the Taylor
expansion in Eq.~(\ref{yexp2}).}
\begin{equation}
A_\perp\simeq \pm\frac{A_E}{\sqrt{1-E/E_f}},
\end{equation}
showing the characteristic $1/\sqrt{1-E/E_f}$ divergence and with the
coefficient $A_E$ being
\begin{equation}
A_E\equiv \left[ 2y_f''E_f\left.\frac{{\rm d}y_f}{{\rm
d}E}\right|_{E_f}\right]^{-1/2}=\left[
\frac{4}{3}\left(\frac{E_\ast}{E_f}-1\right)
\left( 1+2\frac{E_\ast}{E_f}\right)\right]^{-1/2}.
\label{ae}
\end{equation}
It is important to notice that the coefficient $A_E$ depends on how
fast the magnification changes with the observation direction
($y_f''^{-1}$) and also on how fast the fold position moves with
energy (given by d$y_f/{\rm d}E$). 

To find out what are the possible observable effects  of these
magnification peaks, it is useful to introduce the amplification
integrated in an energy bin around $E_f$ (e.g. between $0.9 E_f$ and
$E_f$), since this would be indicative of the potentially observable
signals in a realistic experiment once its finite energy resolution is
taken into account. This has the further advantage that the integrated
magnification becomes finite. Hence, we define
\begin{equation}
A_{int}\equiv\frac{1}{\Delta E}\int_{E_f-\Delta E}^{E_f}{\rm d} E\
\sum_{k=1,2} |A^{(k)}_\perp(E)|\simeq 12.6 A_E\sqrt{\frac{E_f}{10\Delta E}}.
\end{equation}

\FIGURE{\epsfig{file=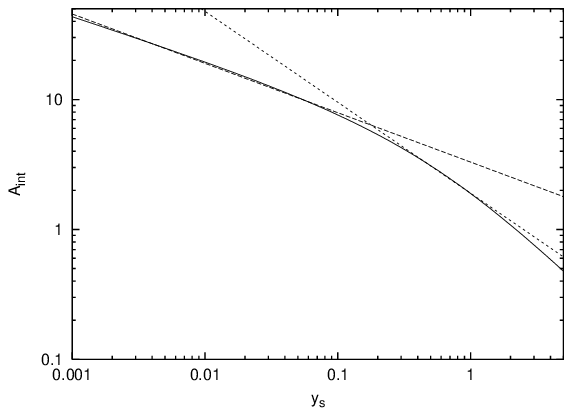,width=9truecm}
\caption{Integrated amplification $A_{int}$ in an energy bin of 10~\%
of $E_f$ as a function of the source position $y_s$ (solid line). 
The dashed lines correspond to
the fits described in the text.}
\label{aint.fig}}

In Figure~\ref{aint.fig} we plot the integrated amplification $A_{int}$
 as a function of $y_s$. The decrease in $A_{int}$ for increasing
 values of $y_s$ is mainly due to the fact that the peaks become
 increasingly narrower for smaller $E_f/E_\ast$.  The  long dashed curve
 corresponds to $A_{int}=3.3 y_s^{-0.38}$, and provides a reasonable fit
 for $y_s<0.2$, while the short dashed curve corresponds to
 $A_{int}=1.9 y_s^{-0.7}$, fitting the results for $0.2<y_s<3$. From
 this we can obtain 
 the probability that the integrated  amplification $A_{int}$
exceeds a given value $A_0$, which equals the probability that $y_s$ be smaller
than the value $y_0$ obtained from the condition that
 $A_{int}=A_0$. Using the previous fits we then get (for $\Delta E=0.1E_f$)
\begin{equation}
P(A_{int}>A_0)=P(y_s<y_0)=y_0\simeq\left\{\matrix{23.1A_0^{-2.63}\ \ A_0>6\cr
2.5A_0^{-1.43}\ \ A_0<6}\right.
\end{equation}
This results imply that there is for instance a probability of $\sim
50$\% that the secondary images lead to an integrated amplification
$A_{int}$ larger than three, while $A_{int}>7$ with $\sim 10$\%
probability. Hence, the effect of these peaks can be in
principle quite relevant.

It is important at this point to consider what would be the
implications of having in addition to the random magnetic fields also
a regular one, coherent over a scale $L_{reg}\gg L_c$.  If both
components have comparable strengths, as happens in the Galaxy,
magnification effects associated to the random one will manifest at
higher energies than those of the regular one. This is because the
lensing depends essentially on field gradients, which are enhanced
when the field variations occur on smaller scales. However, the
possible lensing signatures produced by the random field will be
affected by the presence of the regular one. The most important effect
will be related to the way in which the regular field changes the
``motion'' of the folds with decreasing energies, i.e. the factor
d$y_f/{\rm d}E$ in Eq.~(\ref{ae}).  The motion of the fold is
essentially due to the change in the typical deflections with energy,
and hence for purely random fields it is given by $|{\rm d}y_f/{\rm
d}E|\sim |{\rm d}\delta/{\rm d}E|\simeq \delta/E$. On the other hand,
in the presence of a regular field, the typical deflections will be
given by
\begin{equation}
\Delta\theta\simeq 0.1\frac{B_\perp}{\mu{\rm G}}\frac{L_{reg}}{\rm
kpc}
\frac{10\ {\rm EeV}}{E/Z},
\end{equation}
with $ B_\perp$ being the typical strength of the regular field in the
direction orthogonal to the CR trajectory.
~From Eq.~(\ref{ae}) we then see that the magnification peaks
associated to new image pairs should become narrower, with their
integrated effect being suppressed by the factor
\begin{equation}
\sqrt{\frac{\delta}{\Delta\theta}}\simeq\left(0.65\frac{B_{rms}}{B_\perp}
\frac{L}{L_{reg}}\sqrt{\frac{L_c}{L}}\right)^{1/2},
\end{equation}
where $L\geq L_{reg}$ is the distance traversed across the random field.
For typical galactic parameters (i.e. $B_\perp \simeq B_{rms}/2$,
$L_{reg}\simeq 3$~kpc and $L_c\simeq 100$~pc), this gives
\begin{equation}
\sqrt{\frac{\delta}{\Delta\theta}}\simeq 0.28\left(\frac{L}{3\ {\rm
kpc}}\right)^{1/4}.
\end{equation}
Let us finally mention that something similar happens when we look at
the peaks associated to the short wavelength modes of the
turbulence. These modes can in principle produce peaks on very small
angular scales (and even at somewhat larger energies that the long
wavelength modes), but the motion of the small scale fold will
be determined by the long wavelength modes (or the regular field if
that one is present). This will make the small scale peaks extremelly
narrow and hence less noticeable.

\section{The scintillation regime}
\label{scintillation}

As we have seen, 
the appearance of folds in the mapping from the image plane to the
source plane is
associated to the formation of pairs of new images.
Since for decreasing energies the folds cover an increasing fraction
of the source sky and also new folds are continuously generated, this
means that the
average number of images of a source increases steadily for
diminishing energies. The transition between the regimes where only one
or a few images exist and that in which many images are present 
 may be modelled  with the following
simplified picture, which captures the main processes involved in the
multiple image formation. Let us assume that first at a single energy
$E_1$ a two dimensional rectangular grid of caustics forms, with
separation among them $\Delta\beta\simeq 4L_c/L$. The energy $E_1$
would just be the mean energy of appearance of the folds,
i.e. $\langle E_\ast\rangle$.
 For $E<E_1$, the two folds in each caustic become separated among
them by a typical angular distance $2y_f L_c/L$, with $y_f$ given by
Eq.~(\ref{yfold}) with $E_\ast\to E_1$. 
Since there are two additional images (a total of
three) in the regions covered by a fold, while at the fold
intersections there are $3\times 3=9$ images,  it is then easy to show
that the average number of images for randomly located sources is
\begin{equation}
\langle n \rangle\simeq 1+2 y_f + y_f^2.
\label{n}
\end{equation}
When the folds become  sufficiently wide so that in each face of a
fold there are directions probing uncorrelated magnetic field values,
a second generation of folds can then be generated. This can be
modelled by means of a new network of caustics appearing at an energy
$E_2$ (corresponding typically to the energy at which $y_f\simeq
1$). The width of the new folds will be characterised by a parameter
$y'_f$, given by Eq.~(\ref{yfold}) but with $E_\ast\to E_2$ (with
$y'_f=0$ for $E\ge E_2$). This
would then lead to 
\begin{equation}
\langle n \rangle\simeq(1+2 y_f + y_f^2)(1+2 {y'}_f + {y'}_f^2)\dots \ .
\end{equation}
This process should then repeat itself again at lower energies,
leading to an exponential growth in the mean number of images of the
CR sources. As an example, 
Figure \ref{nmed} shows the numerical results for $\langle n\rangle$ 
and the fits obtained with the
previous expression corresponding to one (long dashes) or two (short
dashes)  generations of folds appearing at
energies  $E_1=85$~EeV (approximately $2~E_c$) and $E_2=20$~EeV, 
with the results being indeed quite satisfactory. 

\FIGURE{\epsfig{file=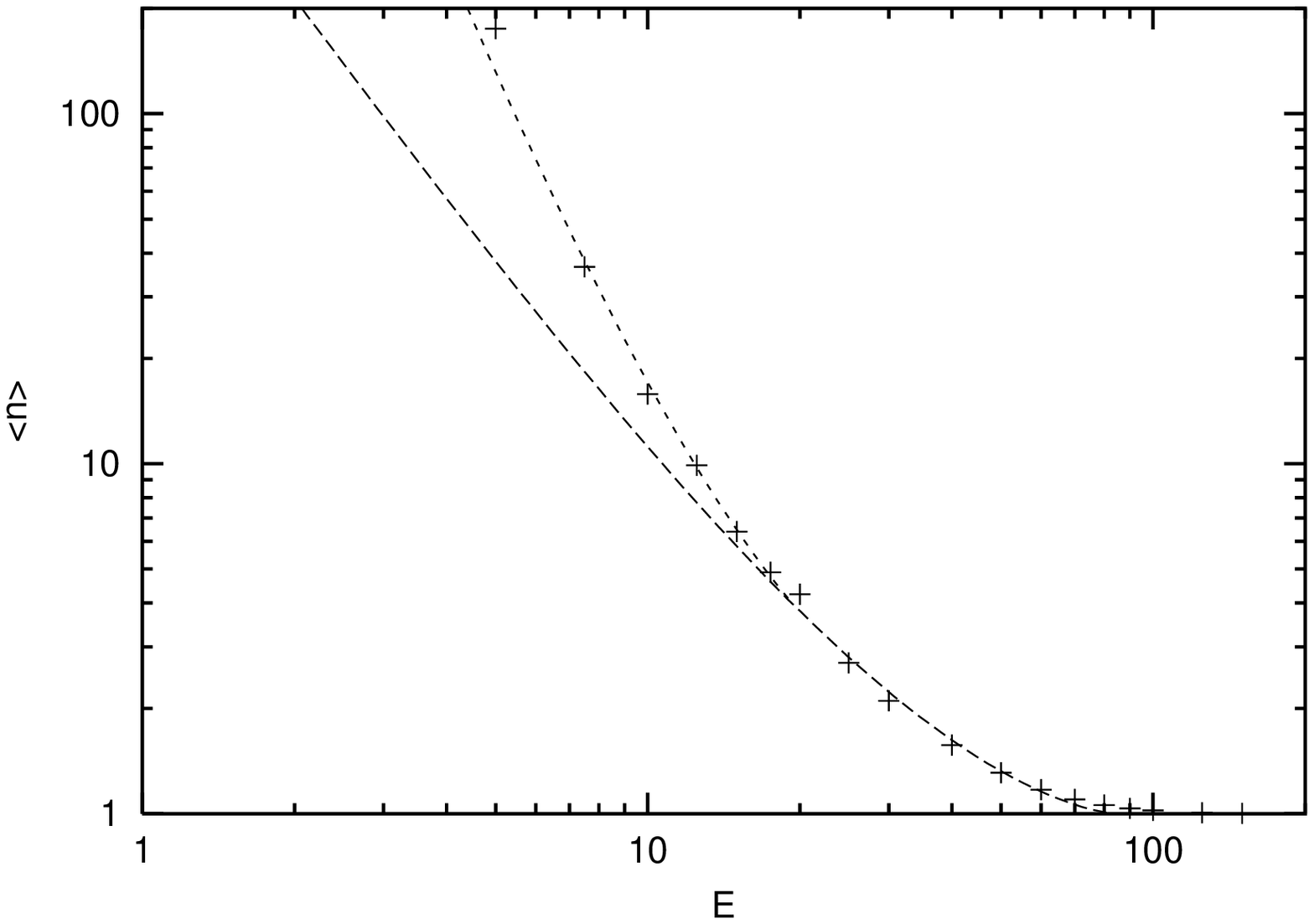,width=9truecm}
\caption{Average number of images of a source vs.  $E$, and
theoretical fits (described in the text). In this example 
$E_c=41~{\rm EeV}$.}
\label{nmed}}

The numerical result for $\langle n \rangle$ in Figure~\ref{nmed} was
obtained using the property that the amplification satisfies 
$A(\theta_1,\theta_2)=J^{-1}$,
where $J$ is the Jacobian of the mapping between the observer's
($\theta_i$) and source's ($\beta_i$) coordinates. Hence, one has that 
\begin{equation}
\frac{1}{4\pi}\int {\rm d}^2\theta A^{-1}=\frac{1}{4\pi}\sum_{images}\int
{\rm d}^2\beta=\langle n\rangle,
\label{nmedeq}
\end{equation}
i.e. that the average of the inverse magnification in the observer's
plane is just the average of the image number in the source plane, and
the first can be obtained computing the magnifications for a dense
grid of directions isotropically distributed around the observer.
The theoretical expression obtained in Eq.~(\ref{nmed1}) reproduces these
results to better than 20\% for $E>E_c/5$, below which the appearance
of multiple folds makes the assumption of Gaussian distributions for
$\kappa$ and $\gamma$ certainly no longer valid.

We can also use this simplified picture of a first network of caustics
forming at an energy $E_1$ to estimate the fraction of the sky in 
which sources have multiple images, as a function of energy.
With arguments similar to those that lead to Eq.~(\ref{n}), 
we estimate $f_s\approx y_f-y_f^2/4$. Using $E_1=85~{\rm EeV}$
we get that more than 20\% of the sky has multiple images around $E=E_c$,
and practically all the sky has multiple images already at
energies between $E_c/3$ and $E_c/4$, as already found before by
different means.

Figure \ref{nmed} shows that already for $E=E_c/3$ there is an
average number of images $\langle n\rangle\simeq 8$, and that 
for $E=E_c/5$ this number has increased to $\langle n\rangle\simeq
30$. This large number of images is due to the continuous creation of
image pairs, which appear largely magnified but the width of the peaks
become increasingly narrow as the energy diminishes, and hence their
integrated effect is reduced. As a result one reaches a regime having
many demagnified (i.e. with $A_{int}<1$) images of every source. These
images will be spread over a typical angular scale $\sim \delta_{rms}$,
which is not necessarily large if $L_c\ll L$. Furthermore, the total
magnification of the many images (averaged in energy bins) will become of
order unity in this regime. This can be understood from the property
that (we are considering here an observer at the center of a spherical
region filled with random fields of constant $B_{rms}$)
\begin{equation}
\frac{1}{4\pi}\int {\rm d}^2\theta =1=\frac{1}{4\pi}\sum_{images}\int
{\rm d}^2\beta A_i.
\end{equation}
If one is in the regime in which all source directions have associated
a large number of images,  one may consider that all points in the source plane
are essentially equivalent, and hence this leads to $\langle
\sum_{images}A_i\rangle\simeq 1$.

The properties of this `scintillating' regime, which in some respects
is reminiscent of the twinkling of the stars produced by the
atmospheric turbulence, are illustrated in Figures~\ref{srshoot}
and \ref{irshoot},
where we present the results of a (cosmic)ray-shooting.  In this
simulation a large number of anti-particles were thrown isotropically 
from the `detector', and those that after traversing a distance $L$ 
point to a direction closer than 1/3 of a degree from that to a 
fixed `source' were recorded. The ratio of this number
to the one that would have been obtained in the absence of magnetic
deflections is just the corresponding magnification. This was repeated
for different energies and the results are plotted in Figure~\ref{srshoot}. 
Superimposed in the same figure are the magnifications of the 
principal image of the source (the one visible at the highest energies)
and of the first few pairs of secondary images. These were evaluated 
numerically with the method described in Section~\ref{numres}, 
which consists in 
tracking the trajectories of three nearby particles for each image. 

\FIGURE{\epsfig{file=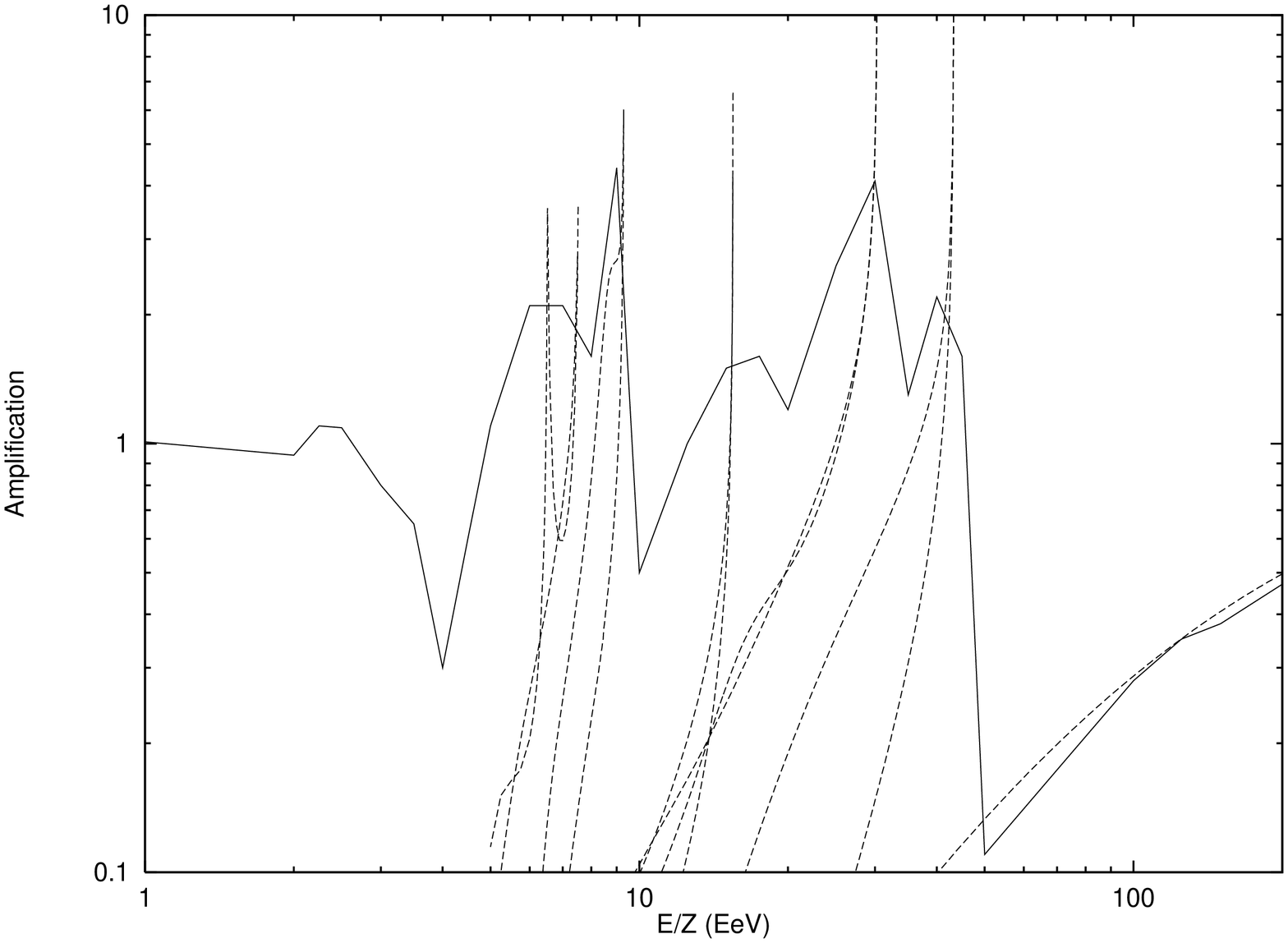,width=8truecm,angle=-90}
\caption{Magnification vs. $E/Z$ of an extended source 
from a ray-shooting simulation (solid line).
Also shown are the magnifications of individual images (dashed lines)
of a point source in the same location.  
The parameters of the simulation are the same as in Figs.~\ref{sheetsM},
\ref{spectra} and \ref{fraccion}, corresponding to $E_c=41~{\rm EeV}$.}
\label{srshoot}}

Notice that in the ray-shooting simulation 
the divergences in the peaks are smoothed by the finite
size of the source. A similar smoothing 
would also result for a point-like source due to the finite energy
resolution of realistic detectors. The peaks associated to
the first few pairs of images are clearly noticeable. As the energy
decreases, the peaks become increasingly suppressed in width, and
there is a progressive transition to the regime with $\langle \sum
A_i\rangle\simeq 1$.

Notice that in this simulation $E_c\simeq 41~$EeV, which is about
the energy at which the first peak is located.  
The source in this example is thus a rather generic one. Sources 
located in about 20\% of the sky should display peaks similar or stronger
than these. The impact of the peaks is generally enhanced by the fact
that the fluxes are usually demagnified above the peak energy. 
Sources in the rest of the sky would display
somewhat narrower  peaks. On the other hand, a smaller fraction of source 
locations would have magnification peaks at higher energies associated
to the principal image, instead of having it demagnified as in the
example of Figure~\ref{srshoot}.

In Figure~\ref{irshoot}  
we show how the different images will look
like at different energies.  Each point in the top panel of  
Figure~\ref{irshoot} represents 1/10 of the unlensed flux of the source,
while in the bottom panels each dot represents 1/100 of it. Before the 
appearance of the first pair of
images the principal one is only slightly displaced (it is assumed to
be at the origin at very high energies) and demagnified (the source is
not very close to the fold formation location).
At lower energies, new magnified images appear and they are spread over the
angular scale $\delta_{rms}\propto1/E$. At the smallest energies
considered, there are very many
demagnified images  but with total magnification close to unity. 

\FIGURE{\epsfig{file=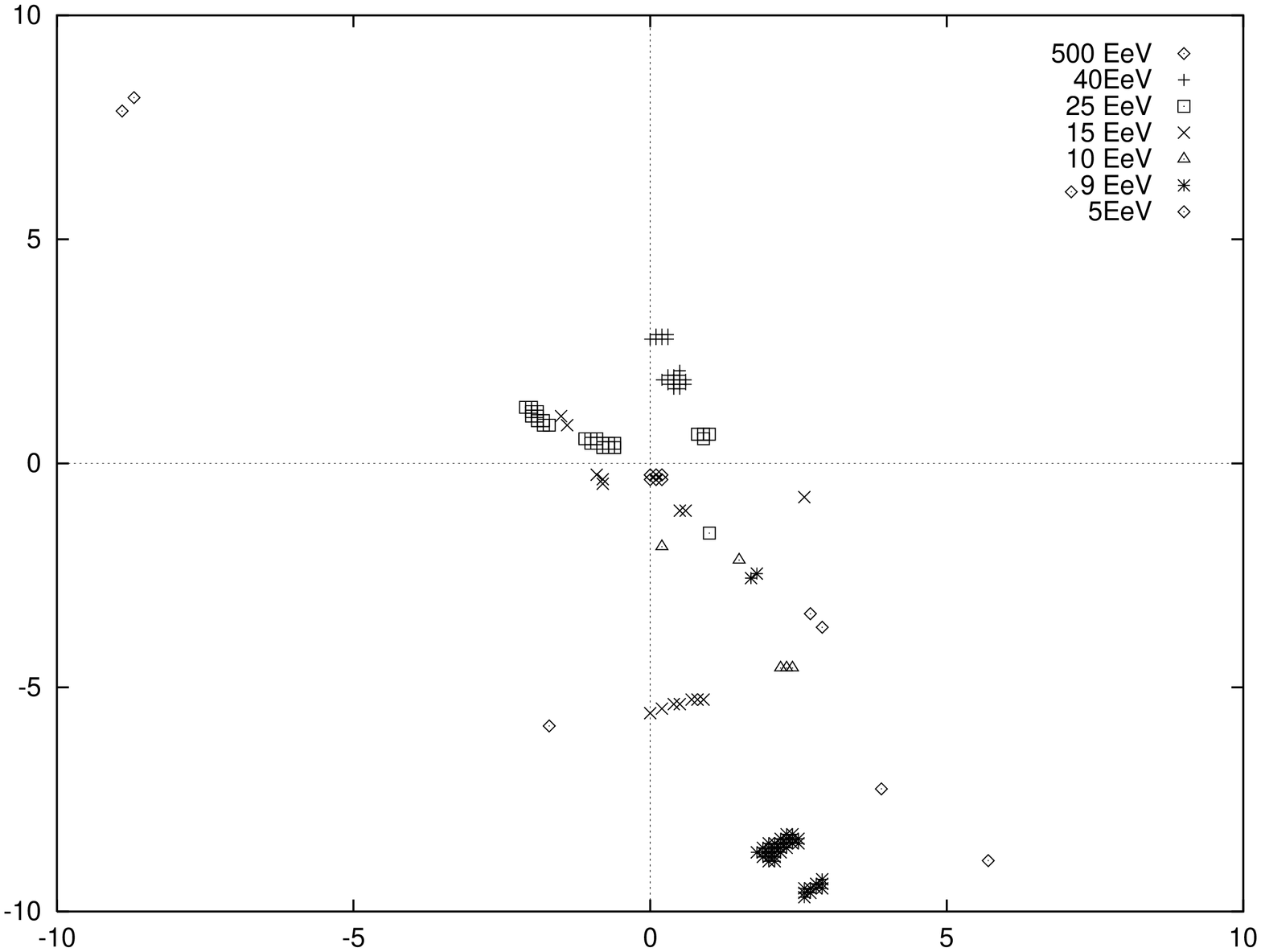,width=9truecm,angle=-90}
{\epsfig{file=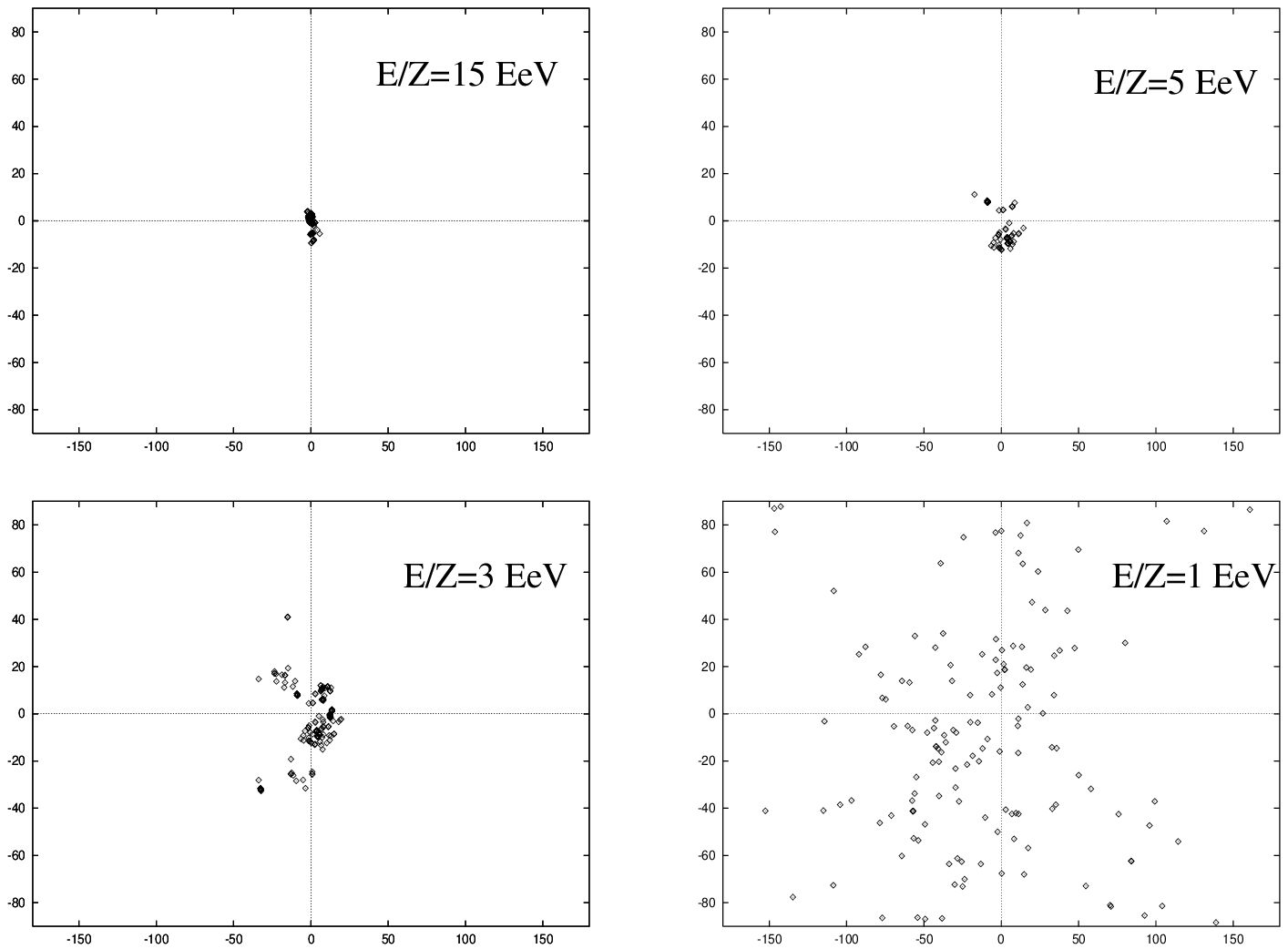,width=5.in}}
\caption{Images of the source obtained from the ray-shooting
simulation corresponding to the previous Figure. Only those with
amplification around and above 1/10 are shown in the top panel,
while in the bottom panels each dot represents 1/100 of the unlensed
flux.}
\label{irshoot}}

Notice that $\delta_{rms}$ becomes of order unity, and then multiple
images of a source cover a significant fraction of the sky, at energies
around and below $E_c/N_c$ (see Eq.~(\ref{Ec})), 
where $N_c\equiv L/L_c$ is the number of 
incoherent domains of the magnetic field traversed in a straigth line
(in our example $E_c=41~$EeV and $N_c=40$).
Spatial diffusion sets in at somewhat lower energies, since the
condition that the CR gyroradius becomes comparable to $L_c$ in a magnetic 
field of strength $B_{rms}$ implies $E\approx E_c/N_c^{3/2}$.

\section{Discussion}
Let us now briefly comment on the possible impact of these results for
the observation of UHECRs. Random magnetic fields are present in the
Galaxy, with few $\mu$G strength and maximum turbulence scale
$L_{max}\simeq 100$~pc, which implies a coherence length $L_c$ somewhere
between 20 and 50 pc, depending on its spectral properties.
Random fields may also  be present on supercluster scales, with much
larger coherence lengths, $L_c\sim $~Mpc, and with strength $\sim
10^{-8}\div 10^{-7}\ $G. In the Galactic case, the typical energy at
which large magnification effects can be present, corresponding to  that
for which $\delta\simeq L_c/L$, is just
\begin{equation}
E_c\simeq Z~41\ {\rm EeV}\frac{B_{rms}}{5\ \mu{\rm G}}
\left (\frac{L}{\rm 2\
kpc}\right)^{3/2}\sqrt{\frac{50~{\rm pc}}{L_c}}.
\end{equation}
This is larger than the typical energy at which lensing effects
associated to the regular galactic magnetic field would appear
($E/Z\simeq 10$~EeV). Furthermore, if the random fields are
concentrated near the galactic plane rather than having a spherical
halo like distribution, CRs arriving at small galactic latitudes will
have associated larger values of $L$, and hence will suffer lensing
effects at higher energies than those arriving from higher latitudes.
The most important signatures of this would be the likely appearance
of significant magnification peaks ($A_{int}>few$) at energies close
to $E_c$ (typically $E_c/3<E_f<2E_c$). These peaks will be associated
with the appearance of the first image pairs, which are very magnified
and appear displaced from the principal image by an angle
$\sim L_c/L$ (but the two new images appear on the same spot in the
sky). This will hence clearly lead to an enhanced signal in a narrow
energy bin (the typical width of the magnification peak, i.e. $\sim
10\%$ of $E_f$), and can then be  a source of clustering of
events. This is similar than what was previously noticed in relation
to the regular fields \cite{ha00}, but manifests at somewhat higher
energies and smaller angular scales (see Eq.~(\ref{deltarms})). 
In the next Section we actually look for the presence of these kind of
signatures in the AGASA data above 40~EeV, and find some significant
hints that could plausibly have their origin in a lensing phenomenon.

For decreasing energies the number of images increases exponentially,
but the lensing effects of each one is suppressed because the
associated peaks in the spectrum become quite narrow. 
In this way one
arrives to the scintillating 
regime, with many images of each source and an overall magnification of
order unity. 
The angular extent of this blurred image of 
the source 
is given by $\delta_{rms}$.  
The presence of
a regular component in the magnetic field would further  suppress the peaks 
produced by the random component.
On the other hand, when for decreasing energies one
reaches the regime where strong lensing effects associated to the
regular field itself are produced, these may be somewhat smoothed by the
finite extension of the large number of subimages just mentioned. 
In any case, one may
understand the effects of the large scale regular fields as resulting
from the folds produced
on a sky which has already been corrugated by the action of the random
field on a much smaller scale.

Regarding the extragalactic random fields, the strong lensing effects
appear at energies 
\begin{equation}
E_c\simeq Z~2\times 10^{20}~{\rm eV}\frac{B_{rms}}{10^{-8}{\rm G}}
\left (\frac{L}{\rm 10\
Mpc}\right)^{3/2}\sqrt{\frac{{\rm Mpc}}{L_c}}.
\end{equation}
which are typically much higher than those associated to the Galactic
fields. These would hence be relevant for supra GZK energies
($E>10^{20}$~eV) even for CR protons.

It should also be noticed that galactic scale random fields present
around extragalactic sources may be relevant in producing an energy
dependent beaming of the fluxes which could also lead 
to interesting features in the observed spectra.

We certainly look forward to the increased CR statistics that will be
accumulated  at ultra high energies in the near future, and which will
allow these effects to be better scrutinized.

\section{Epilogue: hints of lensing in the
AGASA data?}

An interesting signature of the lensing effects, associated to the
high magnification peaks produced when new image pairs appear at the
caustic crossings, is the prediction of an excess of angular
clustering of events with similar energies (actually with similar
rigidities).
As we already noticed in \cite{ha00}, this was strikingly apparent in
the list of doublets and triplets obtained from the combined data of
AGASA, Haverah Park, Volcano Ranch and Yakutsk (eight doublets and two
triplets within $4^\circ$ angular separation) \cite{uc00}. Indeed, most of the
doublets found there are consistent with the two events 
having the same energy, in one of the triplets the energies are in the
ratio $1:2:4$ (as would result e.g. from one event being a proton,
the other a He nucleus and the other a Be nucleus, all with the same
rigidity) 
while in the second
triplet two events have similar energies.

Regarding the analysis of the AGASA data alone \cite{ta99,ta01}, involving five
doublets and one triplet within $2.5^\circ$ angular separation and
$E>4\times 10^{19}$~eV, two of the doublets are consistent with having
the same energy while two events in a triplet also have similar
energies.
To make these statements more quantitative we have studied the
correlations among the AGASA published events both in angle and in
energy, and compared them with simulated events to see the
significance of any excess observed.

Following the studies in \cite{ti01,ta01,al01}, in which one and two
dimensional angular correlations were analyzed to put in evidence the
excess clustering on small angular scales and any possible coherent
deflection produced by a regular component of the galactic magnetic
field, we analysed the correlations of the angular
separation of the events ($\alpha\equiv |\vec\theta_2-\vec\theta_1|$)
and their ratio of energies ($R\equiv max(E_2,E_1)/min(E_2,E_1)$),
where $\vec\theta_{1,2}$ and $E_{1,2}$ are the angular positions and
energies of all possible event pairs chosen among the AGASA (or
simulated) data set.

Taking several bins in $\alpha$ and $R$, we defined the density of
pairs in the bins $f(\Delta\alpha,\Delta R)$ and plotted the
difference between the numbers obtained from the data and the
corresponding averages obtained from a large set of simulated data and
normalizing this to the dispersion in the simulated data, i.e.
\begin{equation}
\rho(\Delta \alpha,\Delta R)\equiv{f_{data}(\Delta \alpha,\Delta R)-
f_{sim}(\Delta \alpha,\Delta R)\over\sqrt{\langle
f^2_{sim}(\Delta \alpha,\Delta R)\rangle-\langle f_{sim}
(\Delta \alpha,\Delta R)\rangle^2}}.
\end{equation}
Each simulation had the same number of events as the real data within
the angular cuts performed. Since the exposure of AGASA is uniform in
right ascension, no cuts were imposed in this angular variable, but
to reduce the sensitivity of the analysis to the unspecified variation
of the exposure with declination (which decreases for increasing
departures from the latitude of Akeno, which is $35^\circ 46'$), we
only considered events with declination in the range
$[0^\circ,70^\circ]$, i.e. essentially within $\pm 35^\circ$ of the location of
the experiment. This leaves 51 out of the 58 published events above
$4\times 10^{19}$~eV.  
We then performed several thousands simulations of sets of 51 events
 distributed randomly within the same angular cuts, with an
energy spectrum d$N/{\rm d}E\propto E^{-2.7}$ and energies above
40~EeV.  

\FIGURE{\epsfig{file=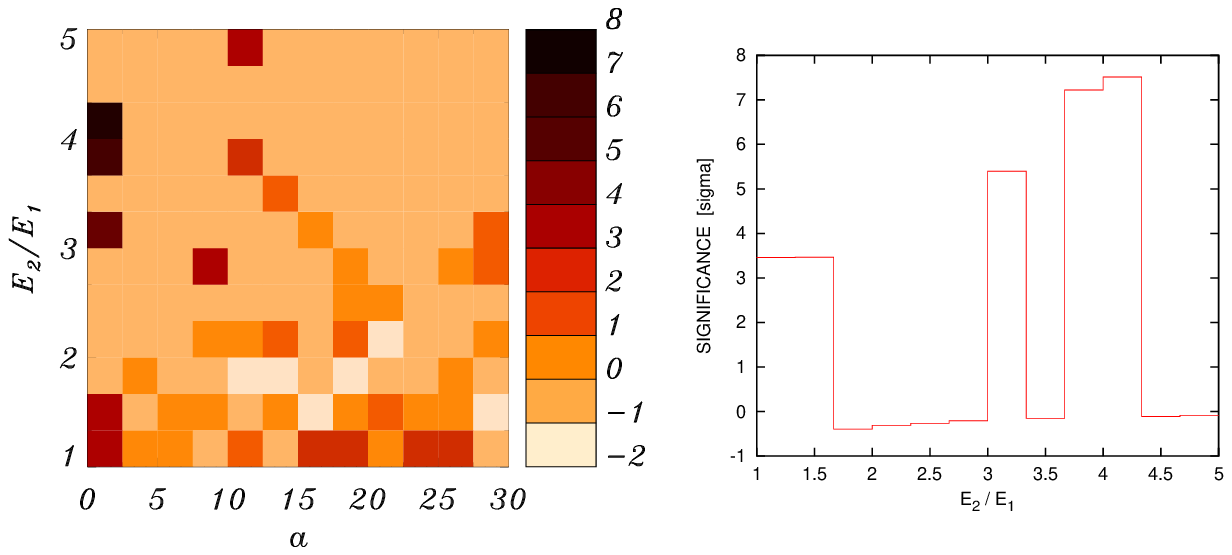,width=15truecm}
\caption{Significance of the correlations in angle and energy ratio 
in the observed AGASA data. The right panel corresponds to the first
angular bin of the left panel ($\alpha<2.5^\circ$).}
\label{corale}}

The results are shown in the left panel of Figure~\ref{corale}. A significant
excess of pairs is found both at small angles and at similar energies.
The right panel shows  the one dimensional correlation in $R$
for pairs separated by less than $2.5^\circ$, which
corresponds to the first angular bin in the left panel.
We see here a more than three sigma excess for events with $E_1\simeq
E_2$, which reflects the fact that two doublets and a pair in the
triplet have this property. The other quite significant peaks
around $R=3$ and 4
reflect the fact that there are doublets with $E_2/E_1\simeq 4.2, 3.7$
and 3.1, something unlikely in a steeply falling spectrum like the one
observed and with the present statistics. 
Magnetic lensing of sources with varied composition is thus
a plausible cause of these peaks. 

Another feature which can be noticed in the observed multiplets is the
fact that all of them have one or two events with energy near or below
$50$~EeV, what could
be indicative of the presence of a threshold energy associated to the
formation of caustics (with the higher energy events in the pairs
corresponding to heavier nuclei).
This fact also explains the excess observed at similar
energies but large angles ($15^\circ\div 25^\circ$) in the two
dimensional correlation (left panel), which seems to be the
result of cross correlation between different doublets having similar
energies.

\bigskip

\acknowledgments

Work partially supported by ANPCYT, CONICET,
and Fundaci\'on Antorchas, Argentina.

\vfill\eject

\end{document}